\documentclass[12pt,document,nofootinbib,onecolumn,preprintnumbers,prd,
balancelastpage]{revtex4-1}
\pdfoutput=1
\usepackage{latexsym}
\usepackage{amssymb}
\usepackage{epsfig,amsmath,graphics}
\usepackage{listings}
\usepackage{color}
\usepackage{dcolumn}
\usepackage{slashed}
\usepackage{cancel}
\usepackage{comment,latexsym} 
\usepackage{bm}
\usepackage{verbatim}
\usepackage{tabularx}
\usepackage{dcolumn}
\definecolor{Mahogany}{rgb}{0.62,0.24,0.15}
\definecolor{colorLink}{rgb}{0.7,0,0}
\definecolor{colorCite}{rgb}{0,.7,0}
\definecolor{colorURL}{rgb}{0,0,0.7}
\definecolor{colorTC}{rgb}{.2,.7,.2}
\definecolor{colorDP}{rgb}{.7,.7,.2}
\usepackage[colorlinks=true,linktocpage=true,linkcolor=black,citecolor=colorCite
,urlcolor=colorURL]{hyperref}
\usepackage{setspace}
\usepackage{xspace}
\usepackage{multirow}
\usepackage{titlesec}
\usepackage[bbgreekl]{mathbbol}
\usepackage{bm}
\usepackage{float}
\usepackage{placeins}
\usepackage{afterpage}

\graphicspath{{Figures/}}

\newcommand{\Lc}{\mathcal{L}}

\def\be{\begin{equation}}
\def\ee{\end{equation}}
\newcommand{\beq}{\begin{equation}}
\newcommand{\eeq}{\end{equation}}
\newcommand{\eref}[1]{Eq.~(\ref{#1})}

\newcommand{\tref}[1]{Table~\ref{#1}}

\newcommand{\fref}[1]{Fig.~\ref{#1}}
\newcommand{\frefs}[2]{Figs.~\ref{#1} and~\ref{#2}}

\newcommand{\lsim}{\!\mathrel{\hbox{\rlap{\lower.55ex \hbox{$\sim$}} \kern-.34em 
\raise.4ex \hbox{$<$}}}}
\newcommand{\gsim}{\!\mathrel{\hbox{\rlap{\lower.55ex \hbox{$\sim$}} \kern-.34em 
\raise.4ex \hbox{$>$}}}}

\newcommand{\GeV}{{\text{ GeV}}}

\newcommand{\gev}{{\text{ GeV}}}
\newcommand{\tev}{{\text{ TeV}}}
\newcommand{\fb}{{\text{ fb}}}

\expandafter\def\expandafter\normalsize\expandafter{%
    \normalsize
    \setlength\abovedisplayskip{8pt}
    \setlength\belowdisplayskip{8pt}
    \setlength\abovedisplayshortskip{8pt}
    \setlength\belowdisplayshortskip{8pt}
}

\titleformat{\section}{\center\normalfont\fontsize{14}{15}\bfseries}{
\thesection.}{1em}{}
\titleformat{\subsubsection}{\center\normalfont\fontsize{12}{15}}{
\thesubsubsection.}{1em}{}

\begin{document}

\title{
 Improving LHC searches for dark photons using lepton-jet
substructure } 

\author{G. Barello}
\affiliation{
Institute of Neuroscience,
Department of Physics, and Institute of Theoretical Science, University of 
Oregon, Eugene, Oregon 97403
\vspace{-4pt}
}

\author{Spencer Chang}
\affiliation{
Department of Physics and Institute of Theoretical Science, University of 
Oregon, Eugene, Oregon 97403
\vspace{-4pt}
}

\author{Christopher A. Newby}
\affiliation{
Department of Physics and Institute of Theoretical Science, University of 
Oregon, Eugene, Oregon 97403
\vspace{-4pt}
}

\author{Bryan Ostdiek$^{\,}$}
\affiliation{
Department of Physics and Institute of Theoretical Science, University of 
Oregon, Eugene, Oregon 97403
\vspace{-4pt}
}

\begin{abstract}
\vskip 40 pt
\begin{center}
{\bf Abstract}
\end{center}
Collider signals of dark photons are an exciting probe for new gauge forces and are characterized by events with boosted lepton jets. Existing techniques are efficient in searching for muonic lepton jets but due to substantial backgrounds have difficulty constraining lepton jets containing only electrons.  This is unfortunate since upcoming intensity frontier experiments are sensitive to dark photon masses which only allow electron decays. Analyzing a recently proposed model of kinetic mixing, with new scalar particles decaying into dark photons, we find that existing techniques for electron jets can be substantially improved. We show that using lepton-jet-substructure variables, in association with a boosted decision tree, improves background rejection, significantly increasing the LHC's reach for dark photons in this region of parameter space.
\vskip -30 pt
$\quad$
\begin{spacing}{1.05}\noindent

\end{spacing}
\end{abstract}

\maketitle
\newpage
\begin{spacing}{1.3}
\pagebreak
%
%

\section{Introduction}
\label{sec:Intro}
High energy physics is at a critical juncture. After the Higgs discovery, the standard model appears to be complete, but has left us in the dark on what, if any, physics lies beyond. Fortunately, in the next few years, data from the LHC's Run 2 have the potential for the most extensive probes into the TeV scale to date, which, if used properly, will  pave the way forward. However, given the uncertainty of what new physics signals will be realized, it is important to cast as wide a net as possible. So in addition to ongoing searches for supersymmetry and other well-studied signals, it is important to consider what signals existing analyses do not address. 

A fruitful direction is to work on improving prospects for  signals that are particularly challenging. For this paper, the general phenomenon we focus on is the dark photon. Nearly a decade ago, dark photons received intense interest due to the cosmic ray anomalies that could be explained by dark matter annihilating into dark photons \cite{ArkaniHamed:2008qn, Pospelov:2007mp}. Dark photons are interesting for their ease in extending the standard model without strong constraints,  since any theory with a massive U(1) gauge boson can kinetically mix with the photon, leading to a predictive framework for the dark photon's couplings to the standard model  \cite{Holdom:1985ag}. This led to a resurgence in fixed-target experiments which are capable of covering a large portion of the unconstrained parameter space (for a review of such experiments, see Ref.~\cite{Essig:2013lka} and for a proposed LHCb search, see Ref.~\cite{Ilten:2016tkc}). As a rough guide, these experiments are most sensitive to dark photon masses between 10 and 200 MeV, which, as we will see, severely impacts collider sensitivities.

At colliders, signals of dark photons are more model dependent, relying on the production of heavier particles that decay or bremsstrahlung into dark photons. For  light dark photons ($\lesssim $ GeV), they decay frequently into charged leptons, leading to collimated leptons, dubbed lepton jets \cite{ArkaniHamed:2008qp, Baumgart:2009tn,Cheung:2009su, Falkowski:2010cm, Falkowski:2010gv, ArkaniHamed:2008qp}. Of the existing lepton-jet searches, muonic lepton jets have much stronger sensitivity due to the smaller rate for hadronic jets to  fake clustered muons. Unfortunately, the fixed-target experiments are sensitive to dark photon masses which kinematically force them to decay to electrons only, the collider background of which are substantial. Collider sensitivity to such dark photons is also limited by splitting analyses into searches for prompt and displaced lepton jets. Since lepton jets get produced in the decays of a heavier particle, a large region of parameter space has the property that many of the lepton jets fall in between the prompt and displaced categories leading to reduced sensitivities. 

In this paper, we address these issues to further optimize the search for electron lepton jets, in order to improve LHC's ability to probe the fixed-target parameter space. To study this, we consider a model we recently proposed for non-Abelian kinetic mixing \cite{Barello:2015bhq}, which has new scalar states that decay into electron lepton jets. A key  virtue of this model is that if a dark photon is discovered at fixed targets, this model predicts these new scalar states have masses within LHC's reach.

In Sec.~\ref{sec:TheModel}, we review the model and its particle content and decays in more detail. In Sec.~\ref{sec:collider}, we consider a benchmark set of parameters to study the collider phenomenology,  which allows us to map LHC sensitivities onto the dark photon parameter space. In Sec.~\ref{sec:Current}, we simulate existing LHC searches to set the current constraints on the benchmark. In Sec.~\ref{sec:collider13}, we estimate the reach for extensions of the existing searches to 13 TeV running. One of our main conclusions is that backgrounds start to become an issue, leading to limited sensitivity gains. In Sec.\ref{sec:Improvements}, we demonstrate that boosted decision trees trained on substructure variables can substantially improve jet rejection while maintaining high lepton-jet efficiencies, leading to a substantially extended sensitivity reach. We conclude in Sec.~\ref{sec:Conclusions}.

\section{Model}
\label{sec:TheModel}

We consider a model that was recently proposed by some of the authors \cite{Barello:2015bhq}, in which kinetic mixing is dominated by the SU$(2)_L$ component of the photon. When a fixed-target experiment discovers the dark photon, such models predict a new particle accessible at the LHC that is charged under both the weak gauge group and the dark U$(1)_D$.  Thus, this is a predictive framework to explore correlated signals at the fixed targets and the LHC.  Since this model was previously explored in Ref.~\cite{Barello:2015bhq}, here we present only the salient features for our study of lepton jets. Despite examining this specific model, we believe our proposed approach to identifying electron lepton jets will be broadly applicable to any model containing them.

To begin, the model adds three new particles: the dark photon $A_D$, the dark Higgs $H_D$, and a scalar mediator $\phi$. The mediator is charged under a U$(1)_D$ gauge symmetry and is also a triplet under the SU$(2)_L$ of the standard model, which generates the kinetic mixing needed for our dark gauge boson to be considered a dark photon. This mixing is controlled by a coupling $\lambda_\text{mix}\, (H^\dag \tau^aH)( \phi^\dag T^a \phi$),  resulting in a kinetic mixing strength,

\begin{equation}
 \epsilon = \frac{g g_D\lambda_\text{mix}}{96\pi^2}\frac{v^2}{m_\phi^2}s_W \sim 
10^{-4}\, g_D\, \lambda_\text{mix} \left(\frac{400 \text{ GeV}}{m_\phi}\right)^2
\end{equation}

\noindent where $g$ is the SU$(2)_L$ gauge coupling, $g_{D}$ is the dark gauge 
coupling, $v$ is the SM Higgs' vacuum expectation value, $m_\phi$ is the mass of $\phi$,  and $s_W$ is the sine of the electroweak 
mixing angle.

After electroweak symmetry breaking, the coupling also induces a mass splitting between the $\phi$ states. The four mass eigenstates are labeled $\chi^\pm$, $\phi^0_{R,I}$, and $\eta^\pm$ (where these are written in descending mass). Their masses are
\begin{equation}
\begin{split}
 m_\chi^2 = m_\phi^2+\frac{\lambda_\text{mix}v^2}{4},  \quad m_\eta^2 = 
m_\phi^2-\frac{\lambda_\text{mix}v^2}{4}.
\end{split}
\label{eq:phi mass splitting}
\end{equation}
and the $\phi^0_{R,I}$ masses are both the bare mass ($m_\phi$). Throughout we will use $\phi$ to refer to all of these states collectively and their individual names when specificity is required. 

As an additional constraint, due to the mass splitting there is a one loop 
contribution to the electroweak precision variable $T$ \cite{Lavoura:1993nq} 
which in the limit of small splitting goes as 
\begin{eqnarray}
T_\text{loop} \sim \frac{\lambda_\text{mix}^2 v^4}{192\pi  s_W^2 c_W^2 m_Z^2 
m_\phi^2} \sim 0.1\, \lambda_\text{mix}^2 \left(\frac{200 \text{ 
GeV}}{m_\phi}\right)^{2}.
\label{eq:T mass splitting}
\end{eqnarray}
where $m_Z$ is the $Z$-boson mass. On the other hand, the contributions to $S$ are negligible, 
so consistency with electroweak precision constraints requires $T< 0.2 
\text{ (95\% C.L.)}$ \cite{Agashe:2014kda}, or equivalently 
\begin{equation}
m_\phi > g_D \left(\frac{10^{-3}}{\epsilon}\right)  140 \text{ GeV} \label{eqn:Tconstraint}.
\end{equation}

\section{Collider phenomenology}
\label{sec:collider}
Looking for correlated collider signals for dark photons motivates certain choices of our signal benchmarks. In particular, parameter values for the model are chosen so that they are relevant for near term fixed-target experiments for dark photons, which are sensitive to kinetic mixing of $O(10^{-5} - 10^{-3})$ and $m_{A_D} \in [10-200]$ MeV. Thus, we will consider  $\epsilon$ in the range of $10^{-5}-10^{-3}$, and we specify to $m_{A_D} = 0.1\gev$. This has important collider ramifications because at this mass the $A_D\rightarrow\mu^+\mu^-$ decay channel is closed. Decays of the dark photon into $\mu$'s have been relatively well studied, see Refs.~\cite{ArkaniHamed:2008qp,Baumgart:2009tn,Cheung:2009su,Falkowski:2010gv, Falkowski:2010cm,Curtin:2014cca, Aad:2015sms,Aad:2015rba, Khachatryan:2015wka}, with stronger sensitivity than lepton jets with only electrons. If our dark photon crossed the $2\mu$ threshold, we would expect extensions of these analyses to maintain strong sensitivity to our model. There are two  potential consequences of these parameter choices. First, given the substantial boosts that the $A_D$ are produced with, we believe going to smaller $A_D$ masses will not significantly change the collider analysis that follows. Second, the smaller range of $\epsilon$'s here leads to displaced decays, which will impact the ability to trigger and reconstruct these events. 

\begin{figure}[t]
\includegraphics[width=0.5\linewidth]{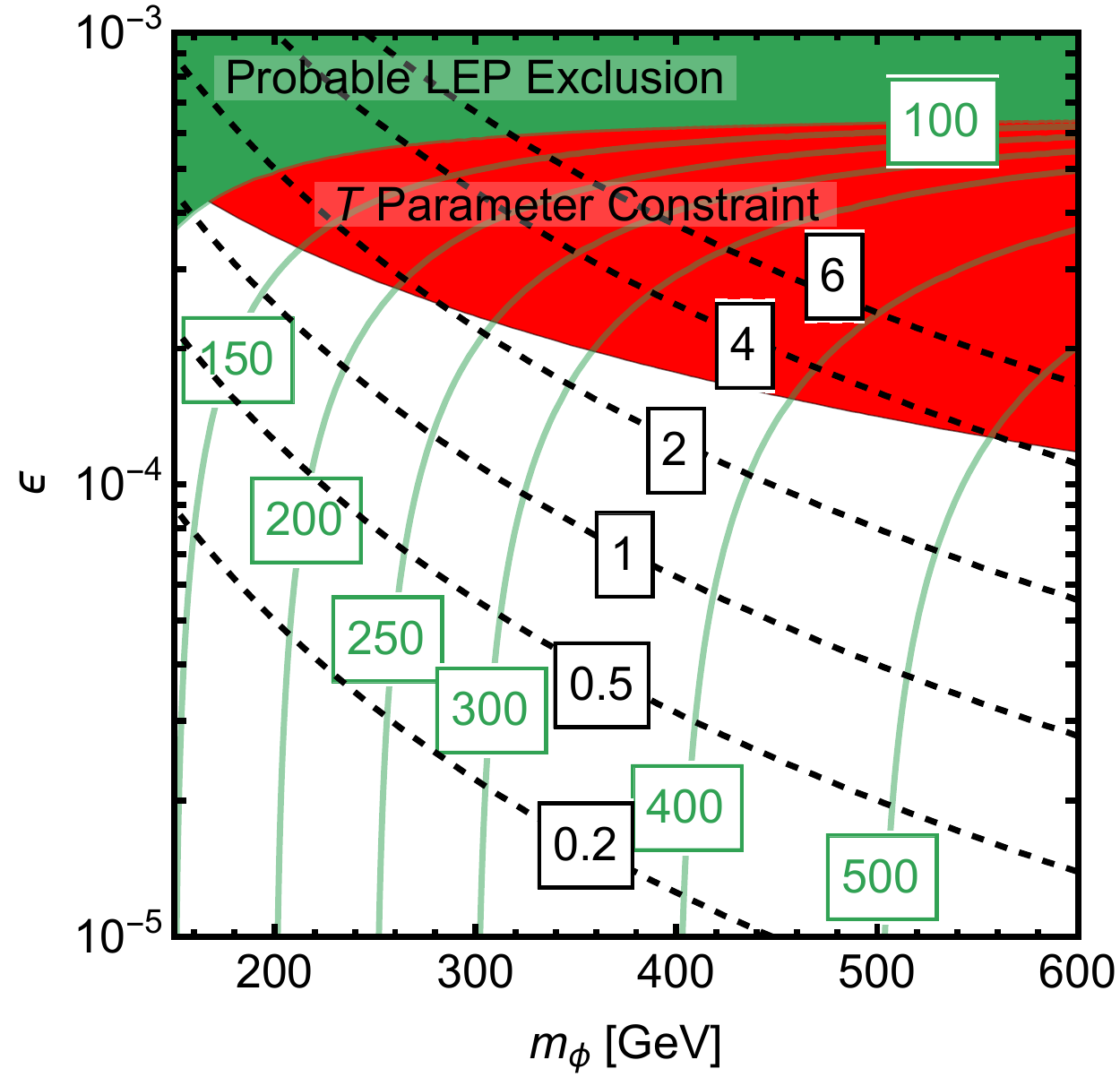}
\caption{Contours of the mixing parameter $\lambda_\text{mix}$ (black dashed) and the mass 
of $\eta$ in GeV (green) for the dark coupling fixed at $g_D = 0.5$. The darker  shaded regions correspond to $m_{\eta} < 100\gev$  which shows the approximate values that LEP naively excluded using searches for charged particles. }
\label{Fig:LambdaAndMEta}
\end{figure}

To pin down the rest of the parameters of our benchmarks,  we choose $g_D = 0.5$, $v_D = 0.2 \gev$, and  $m_{h_D}=0.4\gev$, such that the dark Higgs decays to two dark photons. While examining collider constraints, we scan over a grid of values for $m_{\phi}$ between 150 and 600 GeV, which, along with the $\epsilon$ parameter, determines the value of $\lambda_{\text{mix}}$ at each point. This is plotted in the dashed black contours of Fig.~\ref{Fig:LambdaAndMEta} along with the mass of the lightest triplet state, $\eta^\pm$ (shown as the green contours). Exclusions due to LEP and $T$ parameter considerations are also shown.

The $T$ parameter constraint covers a large swath of the high $\epsilon$ range, but this region should be viewed with some care. First, this assumes no additional precision electroweak contributions exist to cancel or bring one into agreement with the $(S,T)$ combined constraints. Second, if we had chosen a larger value of $g_D$, we see from \eref{eqn:Tconstraint} that the constraint in the $(m_\phi,\epsilon)$ plane weakens considerably, due to the reduced size of $\lambda_\text{mix}$ to achieve a given $\epsilon$. This reduces the splitting between the scalar states, see \eref{eq:phi mass splitting}, and so will nontrivially affect cross sections and efficiencies. However, given the strong exclusion power our collider analyses will place on large $\epsilon$, modifying this benchmark to weaken the $T$ constraint should still have sensitivity in this large $\epsilon$ region.  

The only parameter left undetermined is a coupling, $\kappa$, which governs the decay widths for the $\phi$ particle to decay. The decay rates for $\phi$, mediated by this term, are

\begin{equation}
 \begin{split}
  \Gamma(\phi_R^0\rightarrow hh_D) = \Gamma(\phi_R^0\rightarrow hA_D) &= \frac{\kappa^2v^2}{64\pi m_\phi^3}\left(m_\phi^2-m_h^2\right)\\
  \Gamma(\chi^\pm\rightarrow W^\pm h_D) = \Gamma(\chi^\pm\rightarrow W^\pm A_D) &= \frac{\kappa^2v^2}{128\pi m_\phi^4m_\chi^3}\left(m_\chi^2-m_W^2\right)^3\\
  \Gamma(\eta^\pm\rightarrow W^\pm h_D) = \Gamma(\eta^\pm\rightarrow W^\pm A_D) &= \frac{\kappa^2v^2}{128\pi m_\phi^4m_\eta^3}\left(m_\eta^2-m_W^2\right)^3
\label{eqn:directdecays}
 \end{split}
\end{equation}
as compared to the weak decays of the heavier states
\begin{equation}
  \Gamma(\chi^\pm\rightarrow W^{\pm*}\phi_{R,I}^0) = \Gamma(\phi_{R,I}^0\rightarrow W^{\mp*}\eta^\pm) \approx \sum_{f\bar f'}\frac{N_cG_f^2\Delta m^5}{15\pi^3}
\label{eqn:Wcascade}
\end{equation}
where $G_f$ is the Fermi constant, $\Delta m$ is the mass splitting between $\phi$ states, and $f\bar f'$ includes all fermion pairs except the top-bottom pair for which the splitting $\Delta m$ is too small to produce.

For comparison, for $m_\phi=200\,\text{GeV}$ and $\lambda_\text{mix}=0.5$, the decay rates are
\begin{equation}
 \begin{split} 
  \Gamma(\chi^\pm\rightarrow W^\pm h_D)_\kappa &= 0.63\kappa^2 \GeV \qquad\text{ and }\\
  \Gamma(\chi^\pm\rightarrow W^\pm h_D)_\text{weak} &= 5.3\times10^{-6}\GeV,
 \end{split}
\end{equation}
so for $\kappa\gtrsim3\times 10^{-3}$, the $\kappa$-decays dominate. It is important to note that, even if the weak decays dominated, the decay topology would look very similar in a detector as the $W$-boson from the $\kappa$-decays and the fermions from the weak decays are very soft and so are hard to reconstruct. However, as mentioned in Ref.~\cite{Barello:2015bhq}, if $\kappa$ is small, one notable effect is that half of the time the bottom of the cascade decays, there are two same-sign $\eta$'s, leading to same-sign leptons from the $\eta$-decays.  Although this is an interesting signal, in this paper we will choose a large $\kappa$, so that the direct decays in \eref{eqn:directdecays} dominate over the $W$ cascade decays in \eref{eqn:Wcascade}.

\begin{figure}[t]
\includegraphics[width=0.45\linewidth]{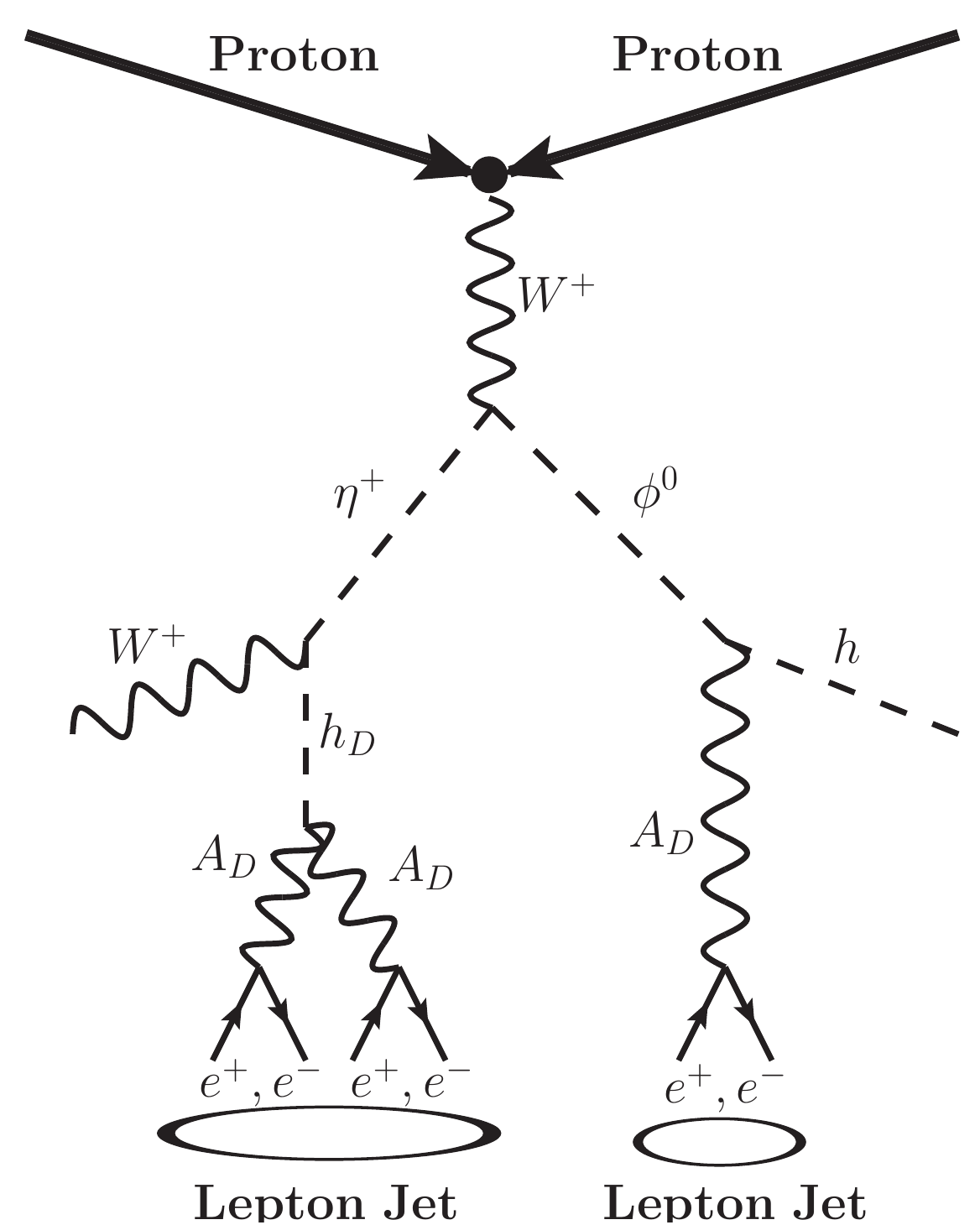}
\caption{Collider production and decay diagram. The dark Higgs/photons are typically very boosted, leading to collimated electron pairs which fail isolation requirements. Special searches are needed to reconstruct these types of objects.}
\label{Fig:typical event}
\end{figure}

\begin{figure}[t]
\includegraphics[width=\columnwidth]{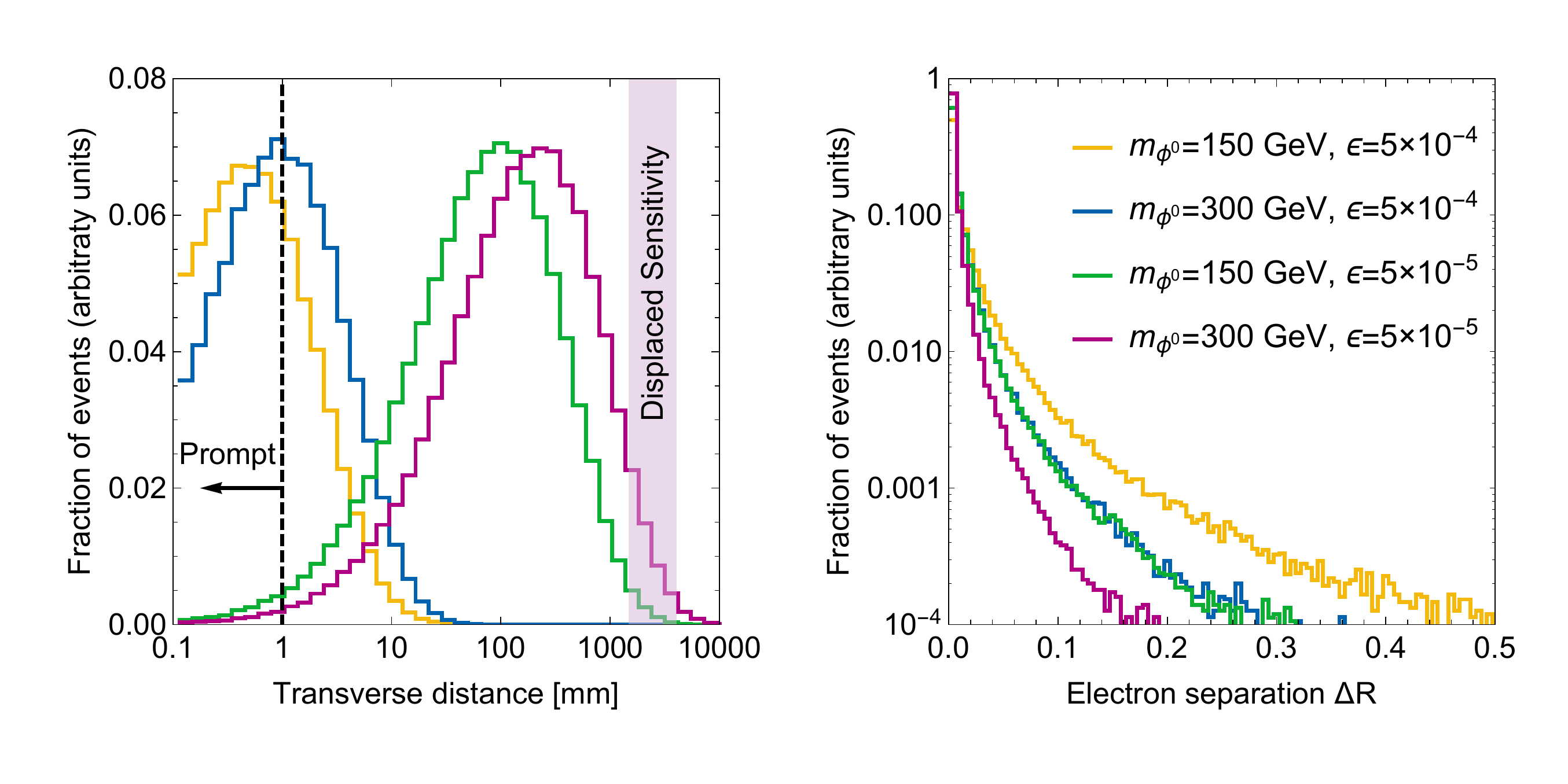}
\caption{Left panel: Decay length for grid points marked above. Right panel: 
Separation of truth-level electrons coming from dark photon decays.}
\label{Fig:DecayLengthandElectronSeparation}
\end{figure}

A typical collider event is shown in \fref{Fig:typical event}. Members of the triplet are pair produced (the lightest have the largest cross section) and decay down to dark Higgs/dark photons along with at least one $W$-boson and either a Higgs or another $W$ depending on the charge of the scalar states produced. As noted by previous collider phenomenological studies of dark photons \cite{ArkaniHamed:2008qp,Baumgart:2009tn,Cheung:2009su,Falkowski:2010gv, Falkowski:2010cm,Aad:2015sms,Aad:2015rba}, the dark Higgs/photons are quite boosted, leading to collimated daughter products. This fact makes searches for dark photons difficult, especially when they can only decay into electrons, because the electrons do not pass isolation requirements. In addition, as the kinetic mixing parameter is decreased, the decay length of the dark photon grows as
\begin{equation}
c \tau_{\gamma_D} = 0.08 \text{ mm} \, \left(\frac{10^{-8}}{\epsilon^2} \right) 
\left( \frac{100 \text{ MeV}}{m_{A_D}} \right)
\label{eqn:dist}
\end{equation}
leading to displaced vertices. \fref{Fig:DecayLengthandElectronSeparation} shows these two effects for four different benchmark values of the kinetic mixing parameter and mass of the triplet. The left panel shows the transverse distance for the production of $e^+e^-$ pairs coming from dark photon decays, while the right panel shows the electrons' separation. The ATLAS search for prompt lepton jets \cite{Aad:2015sms} specifically searches for dark photons which travel less than 1 mm in transverse distance. The purple shaded area marks the region of sensitivity for the ATLAS search for displaced lepton jets \cite{Aad:2015rba, ATLAS:2016jza}. In the next two sections, we examine the constraints on the model coming from existing searches and extrapolate the sensitivity to 300 fb$^{-1}$ of 13 TeV LHC data assuming the same search strategy. 

%
 
\section{Constraints from current search strategies}
\label{sec:Current}

As discussed above, a typical event of pair produced triplets results in two lepton jets along with an assortment of $W$ and Higgs bosons. As the lepton jets can be difficult to use (due to typical isolation requirements), we first checked to see if any limits could be placed on the model using only the $W$-bosons. For instance, if the electroweak decays of $\phi$ dominated the $\kappa$ induced decays, there could be either same-sign leptons. However, searches for same-sign leptons typically target a different type of event, such as $W^+W^+$ production from vector boson fusion (VBF) \cite{Aad:2014zda} or long cascades of supersymmetric models involving large amounts of missing energy and very hard leptons \cite{ATLAS:2013tma}. Our events do not have the very forward jets (as looked for in the VBF search) or the large missing energy, rendering these types of searches inefficient for testing our model. As the lepton jet is a characteristic feature of dark photon models, the strongest limits will come from searches looking for them directly.

There are currently two methods to search for dark photons at colliders: prompt and displaced decay searches. As shown in Eq.~\eqref{eqn:dist} and Fig.~\ref{Fig:DecayLengthandElectronSeparation}, probing lower values of $\epsilon$ requires looking for dark photons which are displaced from the primary vertex. The ATLAS search for displaced lepton jets \cite{Aad:2015rba, ATLAS:2016jza} looks specifically for events which have the dark photon decay within the hadronic calorimeter. This strategy works well if the dark photon is heavy enough to decay to muons, but the backgrounds for electron lepton jets are very large. As such, we find that the best collider limits for this model are currently from the search for prompt lepton jets \cite{Aad:2015sms}, in spite of the stringent 1 mm selection for dark photons that decay promptly. In the remainder of this section, we review this search strategy and recast the search into limits on the model in the $(m_{\phi}, \epsilon)$ plane. For the following results, we used \texttt{FeynRules} \cite{Christensen:2008py} to generate our Lagrangian, \texttt{MadGraph5} \cite{Alwall:2014hca}  to simulate events, \texttt{Pythia6.4} \cite{Sjostrand:2006za} for hadronization and showering, and \texttt{Delphes} \cite{deFavereau:2013fsa} as our detector simulation using the ATLAS detector card.

Before we proceed, note that \texttt{Delphes} does not properly account for dark photons which decay with transverse distance $>$ 1.15 m.  This is because their implementation of the electromagnetic (EM) and hadronic calorimeter has no depth and thus any particles produced farther than 1.15 m are not registered, even if they would have decayed  in the actual ATLAS calorimeters.  The extent of the EM and hadronic calorimeter is out to $4.25$ m.  As you can see from Fig.~\ref{Fig:DecayLengthandElectronSeparation}, this affects the low $\epsilon$ regions, with events that have less calorimeter deposition than in reality.  However, this calorimeter mismodeling should only affect a fraction of events.  In particular, this will have a small effect on our reanalysis of the prompt lepton-jet search since the 1 mm selection only allows sensitivity to high $\epsilon$, but later we will comment on this effect on our proposed modification of the search.  Finally, when we considered the displaced lepton-jet search, due to this \texttt{Delphes} issue, we used the listed efficiencies in Refs.~\cite{Aad:2015rba, ATLAS:2016jza} instead of a detector simulation.  

Following the ATLAS prompt lepton search in Ref.~\cite{Aad:2015sms}, we first use a trigger defined by having an electron tagged track (which need not be isolated) having a $p_T > 60\gev$. Once this requirement is met, the algorithm to define a lepton-jet 
object is implemented. To do this, all tracks with $p_{T} > 5 \GeV$, $|\eta| < 2.5$, transverse impact parameter $d_{0} < 1.0\, \mbox{ mm}$, and longitudinal impact parameter $z_{0} < 1.5\, \mbox{mm}$ are collected. The tracks are then clustered into lepton-jet candidates. To do so, the highest $p_{T}$ track forms the seed for the first lepton-jet candidate, and the next highest $p_{T}$ track within $\Delta R < 0.5$ is added to this candidate. This is repeated until there are no more tracks within $\Delta R < 0.5$ of the lepton-jet candidate, and then a new candidate is formed using the highest $p_{T}$ remaining track. Objects containing at least two tracks with $p_{T} > 10 \GeV$, with one of these tracks being tagged as an electron by \texttt{Delphes}, are now defined as lepton jets. We have verified that our reconstruction procedure matches the efficiency of the ATLAS search.

After defining the lepton-jet objects, several cuts are made to reduce the backgrounds. Only two of the cuts are able to be implemented using \texttt{Delphes}. These are the electromagnetic energy fraction and the track isolation, with the cuts defined as:
\begin{itemize}
\item{Electromagnetic energy fraction ($f_{EM}>0.99$) -- The fraction of the lepton jet's total transverse energy deposited in the EM calorimeter. \texttt{Delphes} reconstructs most lepton jets as Jet objects (not electrons) and so we assign the $f_{EM}$ of the closest reconstructed Jet object as computed by \texttt{Delphes} to each lepton jet.}
\item{Track Isolation $< 0.04$ -- This is defined as the ratio of the scalar sum of all tracks with $p_T>1\gev$ within $\Delta R < 0.5$ of a lepton jet, excluding the EM cluster-matched tracks (those tracks associated with the electron and within $\Delta R<0.05$ of those tracks), to the total lepton jet $p_{T}$.}
\end{itemize}
The other cuts require a more detailed simulation than that done by \texttt{Delphes}. These cuts are the fraction of high-threshold TRT hits, the energy of the strip with maximal energy deposit, and the fraction of energy deposited in the third sampling layer of the EM calorimeter. In the ATLAS analysis, it appears that these cuts greatly affect the background but have a minimal impact on the number of signal counts. Therefore, we do not expect them to significantly affect our final lepton-jet reconstruction efficiency. 

After making cuts on the lepton-jet objects, the signal region is defined as events with at least two  lepton jets. The expected background for two lepton jets is $4.4\pm1.3$ events, while six events were observed with 20.3 fb$^{-1}$ of 8 TeV data. The resulting $95\% ~CL_s$ exclusion is 6 events (expected) or 7.5 events (observed). The yellow area in \fref{fig:13TeV_2ELJ} shows the region of parameter space excluded by this search. The smallest value of $\epsilon$ probed is $\sim 8\times10^{-3}$ while the mass reach is only to $\sim200\gev$.\footnote{The $T$ parameter sets stronger limits after $m_{\phi}=200\gev$ for such large values of $\epsilon$. Again, this could be changed with a larger value of the dark coupling constant, $g_D$, which allows for less mass splitting in the triplet sector. We do not expect that this would change the collider bounds significantly.} In the next section, we extrapolate the same search strategy to the $\sqrt{s}=13\tev$ run of the LHC.

\section{Extension to 13 TeV}
\label{sec:collider13}

\label{sec:Background estimation technique}

In order to update the search predictions to the current run of the LHC, as well 
as extend to other search strategies, we need to estimate the size of the 
backgrounds at $\sqrt{s}=13\tev$. This is not as straightforward as many 
analyses, due to the detector-specific variables used to distinguish the signal 
from the background, which are not possible to model with publicly available detector simulations, so our 
strategy will be to determine the rate at which a jet can fake an electron lepton jet. With this effective
fake rate, the backgrounds for various sources and event topologies will be 
examined.

To find the fake rate, we rely heavily on the ATLAS analysis approach \cite{Aad:2015sms}. In 
particular, ATLAS estimates the background for two, prompt lepton jets using the 
data-driven ABCD method for all backgrounds except for $t\bar{t}$ and diboson 
(VV) plus jets, for which they estimate using Monte Carlo simulations. As the 
fake lepton jets come from hadrons being misidentified as leptons (then 
forming the lepton jet), the hadronic multijet accounts for most of the 
background with $V + \text{jets}$ contributing less than $1\%$ in the 
signal region. The background for the two-electron lepton-jet channel is then given by
\begin{equation}
B_{\text{ABCD}} = 2.9 \pm 0.9 \text{ events} \quad \quad \text{and} \quad \quad 
B_{\text{total}} = 4.4\pm1.3  \text{ events}.
\label{eqn:estimation}
\end{equation}

We simulate the multijet, V+jets, VV+jets, and $t\bar{t}$ +jets backgrounds 
using the same method as the signal generation as described in the previous section.
For each sample, we first find the fraction of events which passes the 
basic trigger used by ATLAS--one electron (does not have to be isolated) with 
a transverse momentum of at least 60 GeV and at least two electromagnetic showers with 
transverse energies of $>35$ and $>25\gev$. The events are then binned in 
jet multiplicity.

To obtain the fake rate ($\varepsilon$), we sum over all of the possible ways in 
which jets in the background samples could produce at least two electron lepton jets, as
\begin{equation}
B_{\text{observed}} = \left( \int \Lc \right) \sum_{s} 
\frac{\sigma_s}{\text{MC}_s} \sum_{j=2}^{\text{max multi.}} n_j \sum_{i=2}^{j} 
\begin{pmatrix} j \\ i \end{pmatrix} \varepsilon^{i},
\label{eqn:BackgroundEstimation}
\end{equation}
where the integrated luminosity is $20.3 \fb^{-1}$, $\sigma_s$ is the total 
cross section for the given background, $\text{MC}_s$ is the number of Monte 
Carlo events for the given source, $n_j$ is the number of Monte Carlo events 
with jet multiplicity $j$, and $i$ is the number of jets which end up faking an 
electron lepton jet. The binomial factor accounts for the different ways of selecting which $i$ 
out of the $j$ jets becomes a fake electron lepton jet. Once all of the contributions have been 
summed, this equation is solved for $\varepsilon$ to get the efficiency per jet 
of events which have already passed the trigger to fake an electron lepton jet.   A more realistic fake rate
would take into account the dependence on energy and other variables but would require more
information about the events ATLAS sees passing the cuts, and thus we stick with this crude, but straightforward, approach.  

Motivated by ATLAS's claim that the background (estimated with the ABCD) method 
is dominated by multijet samples, we first use 
Eq.~\eqref{eqn:BackgroundEstimation} with the multijet sample as the only 
source of background and the observed number as 2.9 events. With this, the fake 
rate is
\begin{equation}
\varepsilon|_{\text{Only multijet in ABCD}} = 6.10 \times 10^{-4}.
\end{equation}
However, when this same fake rate is applied to the other background sources, we 
find that the assumption that all jets are equally likely to fake an electron lepton jet is not 
valid. For instance, this predicts 2.1 events for the V+jets 
channel, but this channel is supposed to account for less than $1\%$ of the 
background \cite{Aad:2015sms}, though using this value for the fake rate is 
conservative as it over-estimates the backgrounds, giving a total predicted 
background of 5.6 events as opposed to ATLAS's $4.4\pm1.3$. For an alternative 
approach, with our V+jets background being larger than expected, we can include 
it in obtaining the 2.9 events. This yields
\begin{equation}
\varepsilon|_{\text{Multijet and V+jets in ABCD}} = 4.64 \times 10^{-4}.
\end{equation}
This matching then underestimates the total background. The results of the two 
methods are summarized in \tref{tab:Background8A} and give a sense of the 
uncertainty in our simplistic jet fake approach. 

\begin{table}[t]
\begin{center}
\caption{Background estimates for $\sqrt{s}=8\tev$. The multijet and V+jets 
samples are included in the ABCD data-driven estimation done by ATLAS, which 
gives a predicted 2.9 events. The first column matches this with only the 
multijet source (which is supposed to be very dominant) but overpredicts the 
total background. The second column matches the ABCD method using the multijet and V+jets sources 
but then underpredicts the total background. The last column averages the two efficiencies and is used throughout the remainder of this paper.}
\begin{tabular}{l | c  | c | c }
\hline
\hline
Source &$\varepsilon=6.1\times10^{-4}$&$\varepsilon=4.6\times10^{-4}$ 
&$\varepsilon=5.4\times10^{-4}$\\
\hline
Multijet & 2.9 & 1.7 & 2.3\\
V+jets & 2.1 & 1.2 & 1.5\\
$t\bar{t}$ & 0.57 & 0.33 & 0.44\\
VV+jets & 0.06 & 0.03 & 0.04\\
\hline
Total & 5.6 & 3.3 & 4.4\\
\hline
\hline
\end{tabular}
\label{tab:Background8A}
\end{center}
\end{table}%

\begin{table}[b]
\caption{Background cross section estimations in fb for $\sqrt{s}=13 \tev$.}
\begin{center}
\begin{tabular}{l c}
\hline
\hline
Source & 2 electron lepton jets \\
\hline
Multijet & $4.03 \times 10^{-1}$  \\
V+jets & $2.42 \times 10^{-1}$ \\
$t\bar{t}$ & $8.69 \times 10^{-2}$ \\
VV + jets & $6.10 \times 10^{-3}$ \\
\hline
Total & 0.738 \\
\hline
\hline
\end{tabular}
\label{tab:B13}
\end{center}
\end{table}

\begin{figure}[t]
\includegraphics[width=0.5\linewidth]{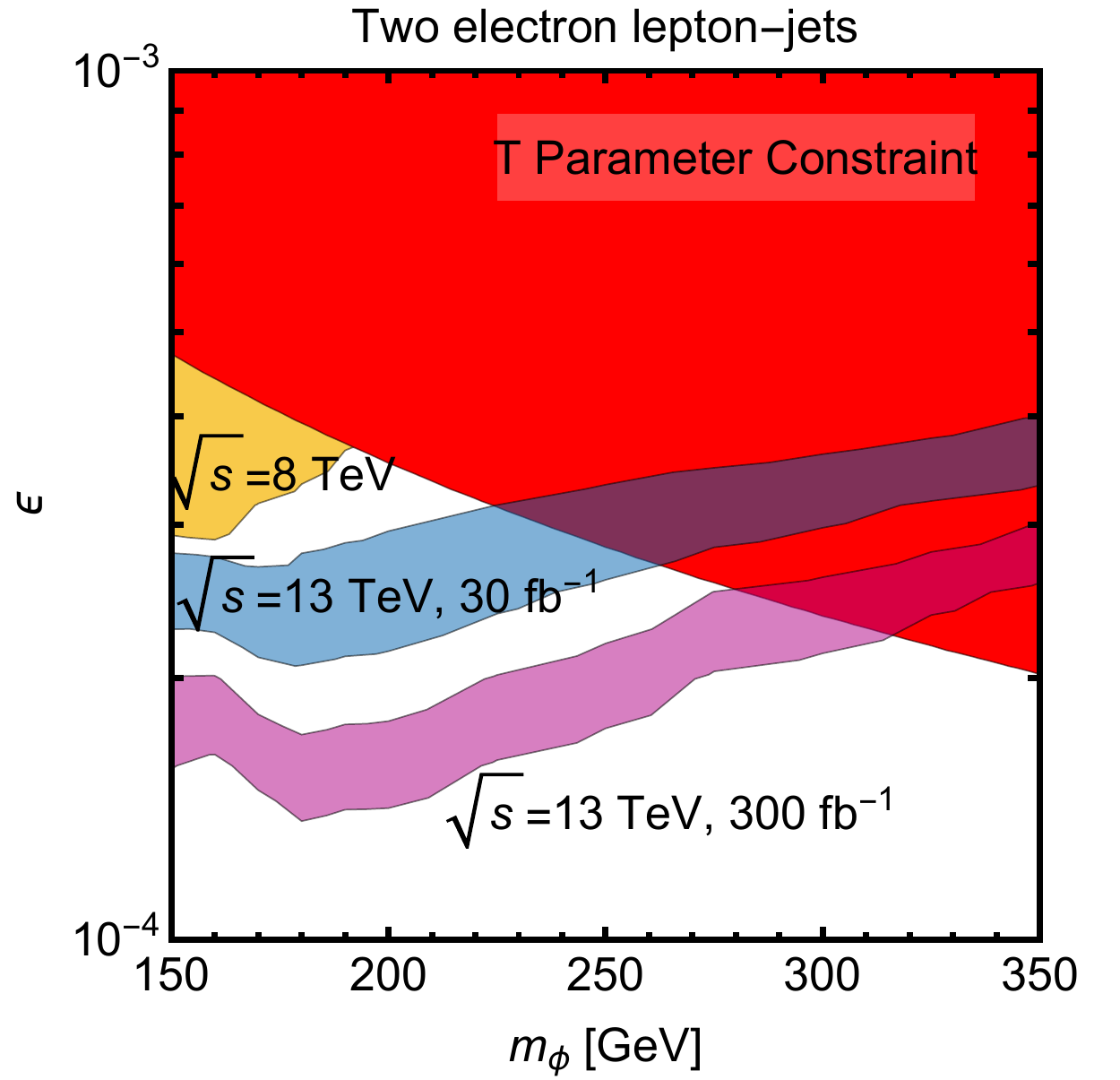}
\caption{The blue (purple) bands show the possible range for $95\%$ C.L. 
exclusions with $30 (300) \fb^{-1}$ of 13 TeV data, allowing the estimated 
backgrounds to float by one-half to twice the extrapolated value. Anything above the 
band would be excluded. The edge of the exclusion limit is predicted to lie 
somewhere within our bands.}
\label{fig:13TeV_2ELJ}
\end{figure}

Extrapolating the backgrounds to $\sqrt{s}=13\tev$ requires some additional 
assumptions. First, we assume that the electron lepton jet is formed with the same algorithm and 
that the trigger is not changed. In addition, we have no information about the 
distributions of the fake electron lepton jets, so again we have to assume a flat efficiency with 
$p_T$ and $\eta$. 
We have also shown that a single jet fake efficiency does not 
translate well between different sources of multijet events, but should give a crude approximation of 13 TeV backgrounds. We will use the average of the fake rates (which happens to 
produce the given total background at $\sqrt{s}=8\tev$) and, to be conservative, give predicted 
exclusions as a band, allowing the background to float between one-half and 
twice the predicted value (which is larger than the uncertainty we found in our 8 TeV background numbers of \tref{tab:Background8A}). \tref{tab:B13} shows the expected size of the 13 TeV backgrounds. With these values, we project the exclusion and discovery 
potential of repeating the same search as done at 8 TeV. This is shown in 
Fig.~\ref{fig:13TeV_2ELJ}. The blue bands show the possible range for the $95\%$ 
exclusion region 
based solely on looking for events with two electron lepton jets after 30$\fb^{-1}$ of 13 TeV 
data. The purple band shows the reach after $300 \fb^{-1}$. The interpretation of these bands is that the exclusion line should fall somewhere within the band, while everything above the line would be ruled out. As noted earlier, the $T$ parameter constraint can be weakened by increasing the benchmark's $g_D$ coupling; however, this will raise the mass of the lightest charged state, which can change the signal cross sections and efficiencies. However, we do expect that these collider searches would have a similar reach if we made such a modification. 

To summarize, the increase in 
cross section from 8 TeV to 13 TeV shows that a larger range of parameter space 
can be covered just by extending the existing analysis.  However, given the projected background cross sections in \tref{tab:B13}, the reach is modest since the background is no longer under control.  In fact, note the restricted range of $\epsilon$, 
which shows that only a small fraction of the fixed-target parameter space ($10^{-3}-10^{-5}$) is coverable 
with $300 \fb^{-1}$.

\section{Improving the search strategy for electron lepton jets}
\label{sec:Improvements}

In the last section, we estimated the reach of the prompt electron lepton-jet search using established methods. While the projected limits are better for the increased energy and luminosity, this search strategy cannot hope to do much better. One of the main reasons for this is in the definition of the lepton jet itself, which requires that the dark photons decay promptly. As shown in Fig.~\ref{Fig:DecayLengthandElectronSeparation}, as $\epsilon$ is further decreased, the dark photon decay length becomes too large to be considered prompt, greatly decreasing the search efficacy. To efficiently probe the whole region of interest, it is crucial to utilize displaced tracks as far as is reliable.  In addition, there may also be other useful variables that define an electron lepton jet beyond the electromagnetic fraction and internal track isolation that can improve our signal acceptance and background rejection.

To examine whether or not these extra features are important, we closely follow the examples of two papers by Ellis, Roy, and Scholtz (ERS) \cite{Ellis:2012sd,Ellis:2012zp}, which developed a technique to look for a similar object: tightly clustered photons (a photon jet).  Please refer to these original references for more details on this approach. The main idea is to use jet-substructure techniques, which are often used for heavy, boosted objects, to classify a reconstructed object as either a QCD jet or a jet with internal structure (like a boosted $W$, top, or photon jet). While the electrons from the dark photon decay are too close to count as isolated objects, when they are clustered into a jet object, we find that there are many features to distinguish the lepton jet from a QCD jet. This process also benefits from an altogether different reconstruction method.

In the current prompt lepton-jet searches \cite{Aad:2015sms}, the lepton jets are put together by merging track objects. However, the strategy presented here follows a more typical jet reconstruction process. First, we take all of the tracks which have a $p_T > 5\gev$ (as identified by \texttt{Delphes}) and make them arbitrarily soft while maintaining their initial direction. We then cluster the tower hits in the electronic and hadronic calorimeters along with these ``new" soft tracks with \textsc{FastJet}\cite{Cacciari:2011ma} using the anti-$k_t$ algorithm \cite{Cacciari:2008gp} with a jet radius of 0.5, only keeping jet objects which have $p_T \ge 20\gev$. By making these tracks arbitrarily soft, they do not affect the overall energy or momentum of the jet.  What they do allow is for us to count the number of charged tracks inside of each jet, $N_{track}$.

We use two slightly different definitions of the $N_{track}$ variable. To begin with, the \texttt{Delphes} tracker is overly optimistic for displaced decays in that the track reconstruction efficiency is the same for any track that appears in their tracker, which does not take into account the loss in efficiency for a track that appears late. To reflect a realistic loss for displaced tracks, we consider two scenarios. In one case, we only use tracks which are prompt [the track starts within 1 mm (radius) of the primary vertex], as required in the ATLAS prompt lepton-jet search. We also study a case assuming that tracks which start within 34 mm of the primary vertex are able to be reconstructed. 
Existing ATLAS searches for displaced electrons \cite{Aad:2015rba} show no dropoff of efficiency to this radius,  and this will allow for a much greater sensitivity for lepton jets with long decay lengths.  For even longer decay lengths in the tracker, one has to also consider backgrounds from converted photons, since boosted dark photons decaying to electrons look similar to converted photons \cite{Agrawal:2015dbf, Dasgupta:2016wxw, Tsai:2016lfg, Bi:2016gca}. At ATLAS, converted photons have to start after the first layer of the tracker, which is at 34 mm. Thus, this case allows us greater acceptance without introducing a new background contribution. 

Continuing with our discussion of jet-substructure variables, for each of the jets we also calculate the fraction of the total jet energy which is deposited in the hadronic calorimeter:
\begin{equation}
\theta_J = \frac{\sum E_{\text{hadronic}}}{E_{\text{total}}}.
\end{equation}
The procedure so far, using the tracks and tower hits, will also pick up objects which can be reconstructed by ATLAS (or \texttt{Delphes}) such as leptons and photons. At this stage, we then veto any jet which lies within $\Delta R < 0.4$ of a \texttt{Delphes}-reconstructed electron, muon, or photon to remove it as consideration as a lepton-jet object.\footnote{A typical dark photon ends up classified as a jet under the standard \texttt{Delphes} ATLAS card.} As a preselection choice, we only keep jet objects which have $N_{track} \ge 1$ (to cut down on photons faking an electron lepton jet) and have $\theta_J < 0.25$ (to cut down on QCD). With these simple cuts, the per QCD jet fake rate is $\sim 3.6\times10^{-3}$.

Jet substructure variables are used on the remaining objects to aid in the classification as a lepton jet or QCD jet. To calculate $N$-subjettiness \cite{Thaler:2010tr, Thaler:2011gf,Stewart:2010tn} for each anti-$k_t$ jet object, the constituents of each jet are reclustered using the $k_T$ algorithm \cite{Ellis:1993tq,Catani:1993hr}, stopping when a total of $N$ subjets are left. $N$-subjettiness is defined as
\begin{equation}
	\tau_N = \frac{\sum_k p_{T_k} \times \text{min} \left\{ \Delta R_{1,k}, \Delta R_{2,k}, \cdots , \Delta R_{N,k} \right\}}{\sum_k p_{T,k} \times R},
\end{equation}
where $k$ runs over all of the calorimeter hits (within the anti-$k_t$ jet), $\Delta R_{i,k}$ is the separation between the $i$th calorimeter cell and the $k$th subjet, and R is the size of the anti-$k_t$ jet (0.5). The utility of $\tau_N$ is that a boosted object with $N$ hard prongs will have a small ratio $\tau_N/\tau_{N-1},$ allowing us to look for the two- and four-prong objects we expect from dark photon and dark Higgs decays. 

After the $N$-subjettiness variables are computed, the constituents are reclustered into five subjets. Of these, only the three hardest are used. This is a version of jet ``grooming" referred to as filtering which helps to remove the effects of pileup \cite{Butterworth:2008iy, Butterworth:2008sd,Butterworth:2008tr}.\footnote{If the anti-$k_t$ jet has less than five constituents, we only cluster until the number of subjets equals the number of constituents.} As the dark photon decays produce highly collimated electrons, we expect most of the $p_T$ of the jet $(p_{T_J})$ to be contained within the leading subjet $(p_{T_L})$. Therefore, we use 
\begin{equation}
\lambda_J = \log\left(1 - \frac{p_{T_L}}{p_{T_J}} \right),
\end{equation}
to help differentiate lepton jets from QCD jets.  The energy-energy correlations of the subjets also show some power to separate the lepton jets from the background. This is defined as
\begin{equation}
\epsilon_J = \frac{1}{E_J^2} \sum_{i>j} E_i E_j,
\end{equation}
where $E_i$ is the energy of the $i$th subjet and $E_J$ is the energy of the whole jet. The sum runs over the three hardest subjets (from the grooming).

Additionally, the spread of the jet is also used. This is defined as 
\begin{equation}
\rho_J = \frac{1}{R} \sum_{i>j} \Delta R_{i,j},
\end{equation}
where $R$ is the jet radius (0.5) and $\Delta R_{i,j}$ is the angular separation of the subjets $i$ and $j$.  Note that each of these three variables, $\lambda_J$, $\epsilon_J$, and $\rho_J$, depend on the clustering algorithm used. We use both the $k_T$ algorithm and the Cambridge/Aachen \cite{Dokshitzer:1997in,Wobisch:1998wt} which ERS found to  increase the separating power.

The last variable we used is the subjet area of the jet, which only makes sense when clustering with a nearest neighbor algorithm (C/A for our purposes). The subjet area fraction is then defined as 
\begin{equation}
\delta_J = \frac{1}{A_J} \sum_{i} A_i,
\end{equation}
where $A_i$ is the area of the $i$th groomed subjet, and $A_J$ is the area of the total jet. Again, see the ERS Refs.~\cite{Ellis:2012zp, Ellis:2012sd} for more information on how these variables are constructed and a discussion on how they are able to distinguish collimated objects from QCD jets.

\subsection{Boosted decision tree}

Using the jet-substructure variables, we now have 13 different properties of each jet object $\left(\theta_J, N_{track}, \log(\tau_1), {\tau_2}/{\tau_1}, {\tau_3}/{\tau_2}, {\tau_4}/{\tau_3}, \left\{ \lambda_J, \epsilon_J, \rho_J \right\}|_{k_T}, \left\{ \lambda_J, \epsilon_J, \rho_J, \delta_J  \right\}|_{C/A}\right)$. Each property shows some separating power, so we use a multivariate analysis to distinguish between the lepton jets and the QCD jets. To do so, we first run through all of the model points in our parameter space to collect each reconstructed jet object (as defined above and using the appropriate track radius) which has a truth-level dark photon within $\Delta R < 0.5$. Next, we collect each of the reconstructed jets from our multijet and $t\overline{t}$ samples to serve as background examples. We then use TMVA \cite{Hocker:2007ht} to train a boosted decision tree (BDT) using AdaBoost \cite{Freund:1997xna} on 400 trees which each have a maximum depth of 3.

The default hyperparameters for AdaBoost in the TMVA setup are used, including splitting the sample in half, using half for training and half for validation. After training the BDT (using only the training sample), the resulting BDT is applied to both the training and validation samples, and the results are found to differ less than $1\%$, showing that overtraining is not occurring. We find that the most important variables are $\theta_J, N_{track}, \lambda_J \text{, and }  {\tau_2}/{\tau_1}$.  The power to separate the signal jets from the background jets is very good for both track distance requirements. The area under the receiver operating characteristic (ROC) curve is 0.912 and  0.961 for 1 and 34 mm track requirements, respectively. Note that the results could probably be improved by tuning the hyperparameters, training on a larger fraction of the samples, or using other MVA techniques. However, this simple setup shows significant gains over the original prompt lepton-jet search.  In comparison, we found a simple cut-based analysis on these top four variables gives signal efficiencies worse than the BDT by around $10\%$ for the 1 mm track requirement and $3\%$ for 34 mm; see the Appendix for more details.  Although the cut-based approach may be simpler to validate and implement into experimental analyses, we will focus on the BDT results in the rest of the paper to maximize the experimental reach.   

\begin{figure}[t]
\includegraphics[width= \columnwidth]{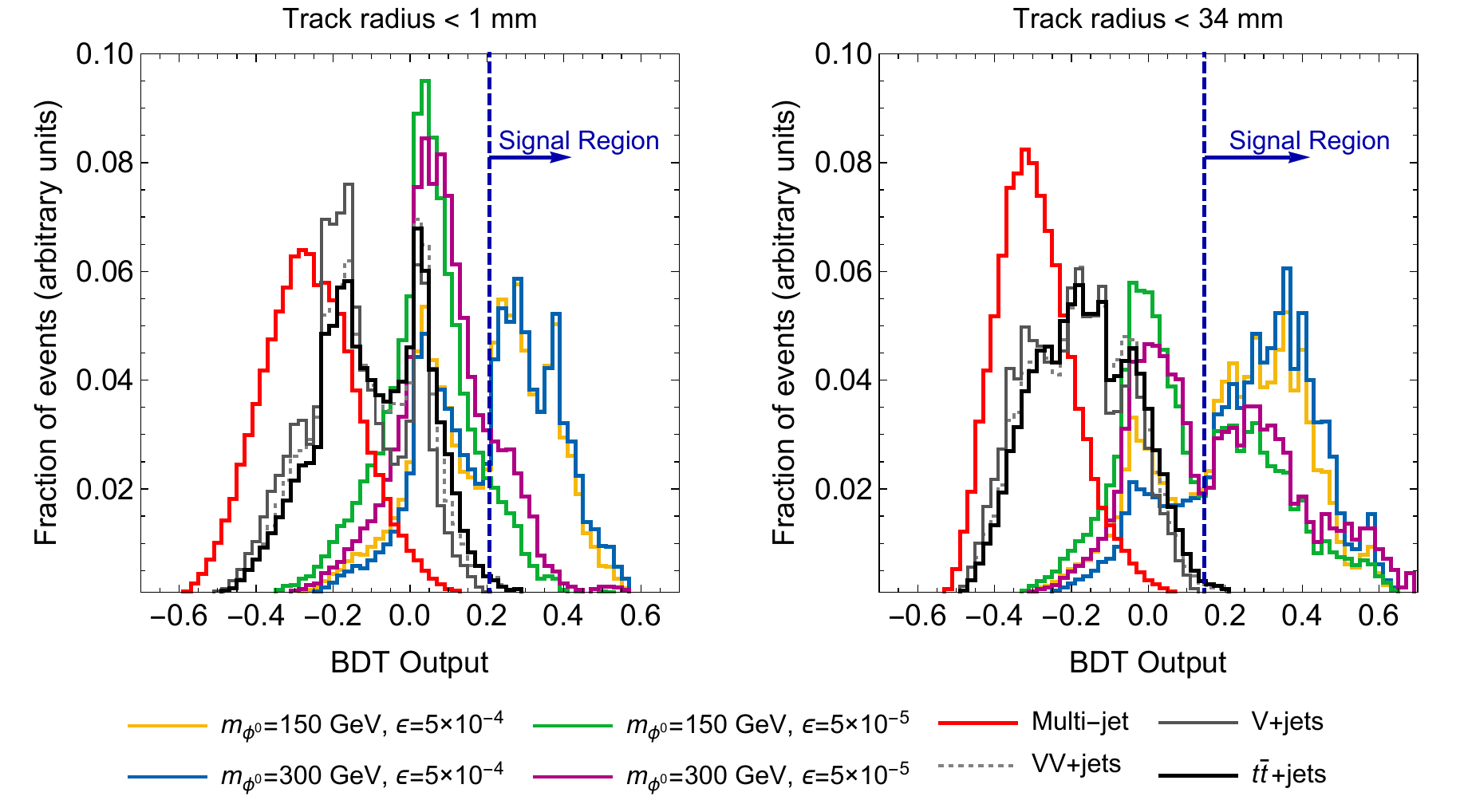}
\caption{Distribution of the backgrounds and benchmark model points after passing through the multivariate analysis. The left (right) panel only includes tracks which start within a radius of 1 mm (34 mm) of the primary vertex. The signal region is defined by the BDT value which yields a fake acceptance rate of 0.01 in the validation sample.}
\label{fig:BDTOut}
\end{figure}

For each of the trees, we select a cut on the BDT output which yields a fake acceptance rate of 0.01. For the 1 mm track requirement, the cut on the BDT is then 0.21, which leads to a signal efficiency of 0.37. Allowing for more displaced tracks increases the acceptance, using a BDT cut of 0.15, leading to a signal efficiency of 0.66. \fref{fig:BDTOut} shows the probability density of the different background sources along with the four benchmark models considered above. Note that the ROC curve numbers, as well as the acceptance rates quoted above are for the training and validation samples, which contain all signal parameter points and only the multijet and $t\overline{t}$ backgrounds.\footnote{We find that training on all model parameter points at the same time performs better than a single model point. This is due to the much larger sample size when using all of the model points, so the BDT is able to more effectively learn what a lepton jet, collectively, looks like.}

In Fig.~\ref{fig:BDTOut}, we see that individual model points and backgrounds behave quite differently under the BDT. There is also an interesting feature in the BDT output (especially for the 1 mm case) resulting in a peak near a value of 0. This bump does not appear in the multijet sample but does for all of the backgrounds which  contain a $W$-boson decaying leptonically. We find that a BDT trained solely upon distinguishing lepton jets from this leptonic $W$ background still contain this extra feature, suggesting that this is something systemic in our BDT approach (e.g. depth of the trees and features in the events) and not on the training samples. Although such a feature makes one worry, such features also appear in ERS \cite{Ellis:2012zp, Ellis:2012sd}, and, as we shall see, our BDT still allows a significant improvement in separating signal from the backgrounds.   

\begin{figure}[t!]
\includegraphics[width=.9\columnwidth]{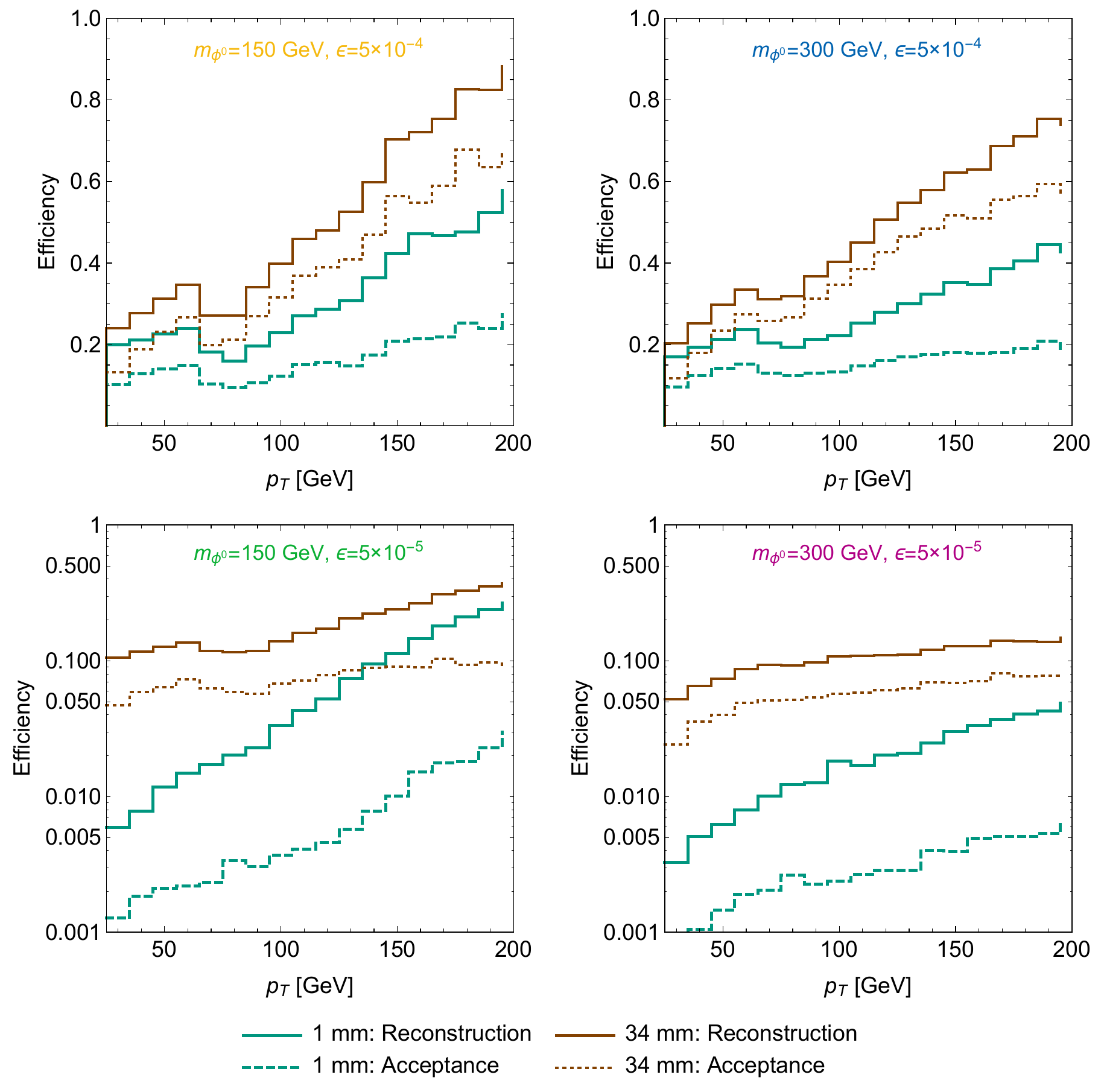}
\caption{Efficiencies for the reconstruction and acceptance of dark photons in the benchmark models. The reconstruction efficiency (sold lines) is defined as the number of reconstructed jet objects which contain exactly one truth-level dark photon within $\Delta R < 0.4$ divided by the total number of dark photons in the same $p_T$ window. Note that, as the reconstructed object may not have the same $p_T$ as the dark photon, the reconstruction rates may be greater than one. The dashed lines show the acceptance rates, defined as the number of jet objects passing the BDT but which have exactly one truth-level dark photon within $\Delta R < 0.4$ divided by the number of truth-level dark photons within the same $p_T$ window. The brown and green lines compare the different track distance requirements; allowing for more displaced tracks greatly improves the acceptance of the models with lower $\epsilon$ and thus longer decay lengths. The four panels display the four different benchmark models considered.}
\label{fig:DPAcceptance}
\end{figure}

To show the benefit of using more displaced tracks, the reconstruction efficiency and the acceptance for single dark photons and dark Higgs are shown  in \frefs{fig:DPAcceptance}{fig:DHAcceptance}, respectively.  For the reconstruction efficiency (solid lines in \frefs{fig:DPAcceptance}{fig:DHAcceptance}), we first find all of the jet objects (as defined above) and bin them according to their transverse momentum. Objects with exactly one (two) truth-level dark photon(s) within $\Delta R < 0.4$ are classified as a \emph{reconstructed} dark photon (dark Higgs) object. 
The reconstruction efficiency is defined as the $p_T$ distribution for the reconstructed dark photon divided by the distribution of the truth-level dark photons, with a similar procedure for the dark Higgs.
With these definitions, it is possible that a truth-level dark object could be reconstructed into a different $p_T$ bin, so the upper bound is not necessarily unity. Next, the acceptance is plotted using the same procedure, but only counting jet objects which also pass the appropriate BDT value cut. At this point, it is important to remember that to be considered as a jet object there must be at least one track clustered in the object. Thus, we see that reconstruction efficiency is very bad for the benchmarks with lower $\epsilon$ values. While the acceptance is not large for the jets with a track-length requirement of less than 34 mm, we will show that this method is still an improvement over the previous search strategy. In addition, we point out that the acceptance for the dark Higgs can be much greater. This is due to there being two dark photons with in the object, increasing the likelihood that one decays in time to be considered.

\begin{figure}[t!]
\includegraphics[width=0.9\columnwidth]{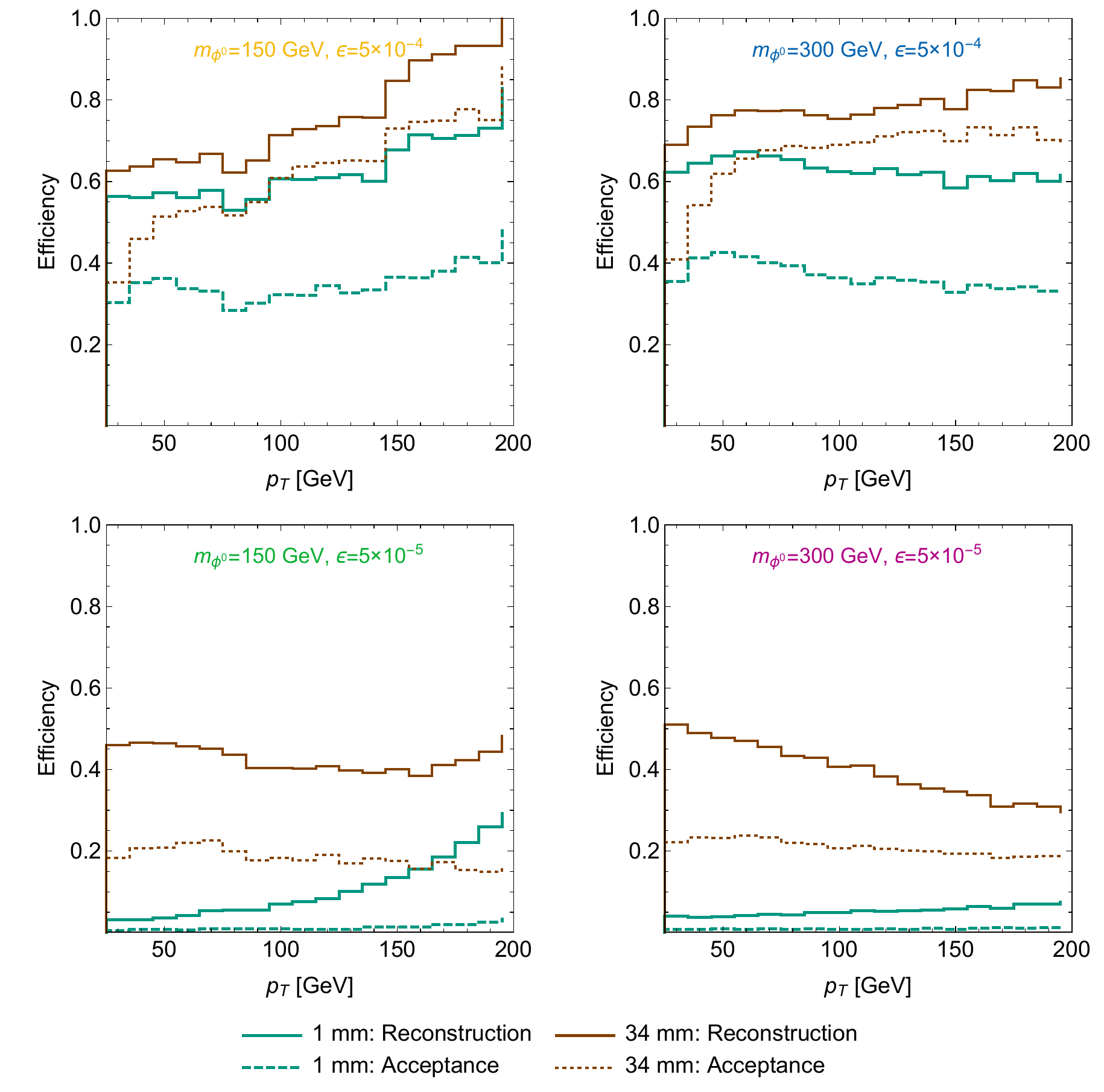}
\caption{Same as previous figure, but now for dark Higgs objects, rather than single dark photons. }
\label{fig:DHAcceptance}
\end{figure}

\subsection{Results using substructure techniques}

In this section, we analyze the results from the boosted decision tree defined in the previous section. First, we determine the per-jet fake rate for each type of background source, including V+jets and VV+jets, which were not included in the training of the BDT, by finding the total number of jets passing the BDT and dividing by the total number of jets in each sample. The results of this are shown in \tref{tab:Backgrounds}. The per-jet fake rate obtained in this method has a large range, depending on the type of background; however, each source has a rate that is lower than our largest estimates of the rate based on the 8 TeV search done by ATLAS shown in \tref{tab:Background8A} \cite{Aad:2015sms}.
For the two different assumptions on the track distance requirements, we have estimated the background cross sections for having a lepton jet along with an isolated lepton and for having two lepton jets. These are summarized in \tref{tab:Backgrounds}. Notice that the background cross sections are larger for the requirement that the tracks start within 34 mm in radius from the primary vertex. This is eventually offset by the greater signal acceptance.  In comparison to our projected backgrounds of the existing prompt lepton-jet search, we see that the total backgrounds have been reduced by almost 2 orders of magnitude.

\begin{table}[t]
\caption{Backgrounds estimated when using jet substructure and the BDT to reconstruct the lepton jets.}
\begin{tabular}{l | c | c | c | c }
\hline
\hline
\multicolumn{5}{c}{Track radius $<1$ mm}\\
\hline
Source & Per-jet fake rate & $\sigma$(1 lepton jet)[fb] & $\sigma$(1 lepton jet + lepton)[fb]  & $\sigma$(2 lepton jets)[fb] \\
\hline
Multijet & 6.68 $\times10^{-6}$ & 6.73 & 1.20$\times10^{-3}$ & 6.16 $\times 10^{-5}$ \\
$t\overline{t}$ & 1.49$\times10^{-4}$ & 21.4 & 12.7 & 6.60$\times10^{-3}$ \\
V+jets & 1.94$\times10^{-5}$& 19.2 & 9.46 & 3.12$\times10^{-4}$ \\
VV +jets & 1.51$\times10^{-4}$ & 2.67 & 1.48 & 4.77$\times10^{-4}$ \\
\hline
\multicolumn{5}{c}{Track radius $<34$ mm}\\
\hline
Source & Per-jet fake rate & $\sigma$(1 lepton jet)[fb] & $\sigma$(1 lepton jet + lepton)[fb]  & $\sigma$(2 lepton jets)[fb] \\
\hline
Multijet & $7.32\times10^{-6}$ & 7.38 & 1.31$\times10^{-3}$ & 7.40 $\times 10^{-5}$ \\
$t\overline{t}$ & 1.91$\times10^{-4}$ & 27.5 & 16.3 & 1.09$\times10^{-2}$ \\
V+jets & 2.72$\times10^{-5}$& 27.0 & 13.3 & 6.16$\times10^{-4}$ \\
VV +jets & 1.15$\times10^{-4}$ & 2.03 & 1.13 & 2.77$\times10^{-4}$ \\
\hline
\hline
\end{tabular}
\label{tab:Backgrounds}
\end{table}

The estimated number of background events are shown in \tref{tab:SI95}, which shows that our BDT search for two electron lepton jets has reasonable backgrounds up to $300$ fb$^{-1}$. Under the assumption that the experiment observes the number of expected events, the number of signal events that can be excluded at the 95$\%$ C.L. is found using
\begin{equation}
\frac{\int \delta b_i \text{Gaus} \left(\delta b_i, \frac{\sigma_{b,i}}{b_i} \right) \times \text{Pois} \left(n_i | b_i (b_i (1+\delta b_i)+ s_{i,95}\right)}{\int \delta b_i \text{Gaus}\left(\delta b_i, \frac{\sigma_{b,i}}{b_i}\right) \times \text{Pois}\left(n_i|b_i(1+\delta b_i) \right)} = 0.05,
\label{eqn:si95}
\end{equation}
where $b_i$ is the expected number of background, $n_i$ is the observed number of events, and $\sigma_i$ is the error \cite{6428calculation,4476calculation}. This is solved numerically to get the value of $s_{i,95}$. For the error, we use the square root of the background. To be conservative, a band is made by computing an exclusion assuming the background is larger or smaller by a factor of 2. 

\begin{table}[t]
\caption{$B$ represents the expected number of background events for a given signature. $S_{i}$ is the number of extra signal events that would be excluded at the $95\%$ C.L. if the experiment observes the expected number of background events, where $i$ is a multiplicative factor for the expected background.}
\begin{tabular}{l | c c c c | c c c c  }
\hline
\hline
& \multicolumn{4}{c|}{$30 \fb^{-1}$} & \multicolumn{4}{c}{$300 \fb^{-1}$} \\
\hline
Track radius $<1$ mm	& $B$ & $S_{1/2}$ & $S_{1}$ & $S_{2}$ &$B$ & $S_{1/2}$ & $S_{1}$ & $S_{2}$ \\
\hline
1 lepton jet + lepton & 709 & 66.7 & 94.2 & 132 & 7092 & 208 & 293 & 413  \\
2 lepton jet & 0.22 & 3.32 & 3.59 & 4.04 & 2.24 & 5.15 & 6.62 & 8.86 \\
\hline
\hline
Track radius $<34$ mm	& $B$ & $S_{1/2}$ & $S_{1}$ & $S_{2}$ &$B$ & $S_{1/2}$ & $S_{1}$ & $S_{2}$ \\
\hline
1 lepton jet + lepton & 922 & 75.9 & 107 & 150 & 9219 & 237 & 334 & 472  \\
2 lepton jet & 0.36 & 3.48 & 3.86 & 4.50 & 3.55 & 6.06 & 8.02 & 10.9 \\
\hline
\hline
\end{tabular}
\label{tab:SI95}
\end{table}

\begin{figure}[t!]
\includegraphics[width=\columnwidth]{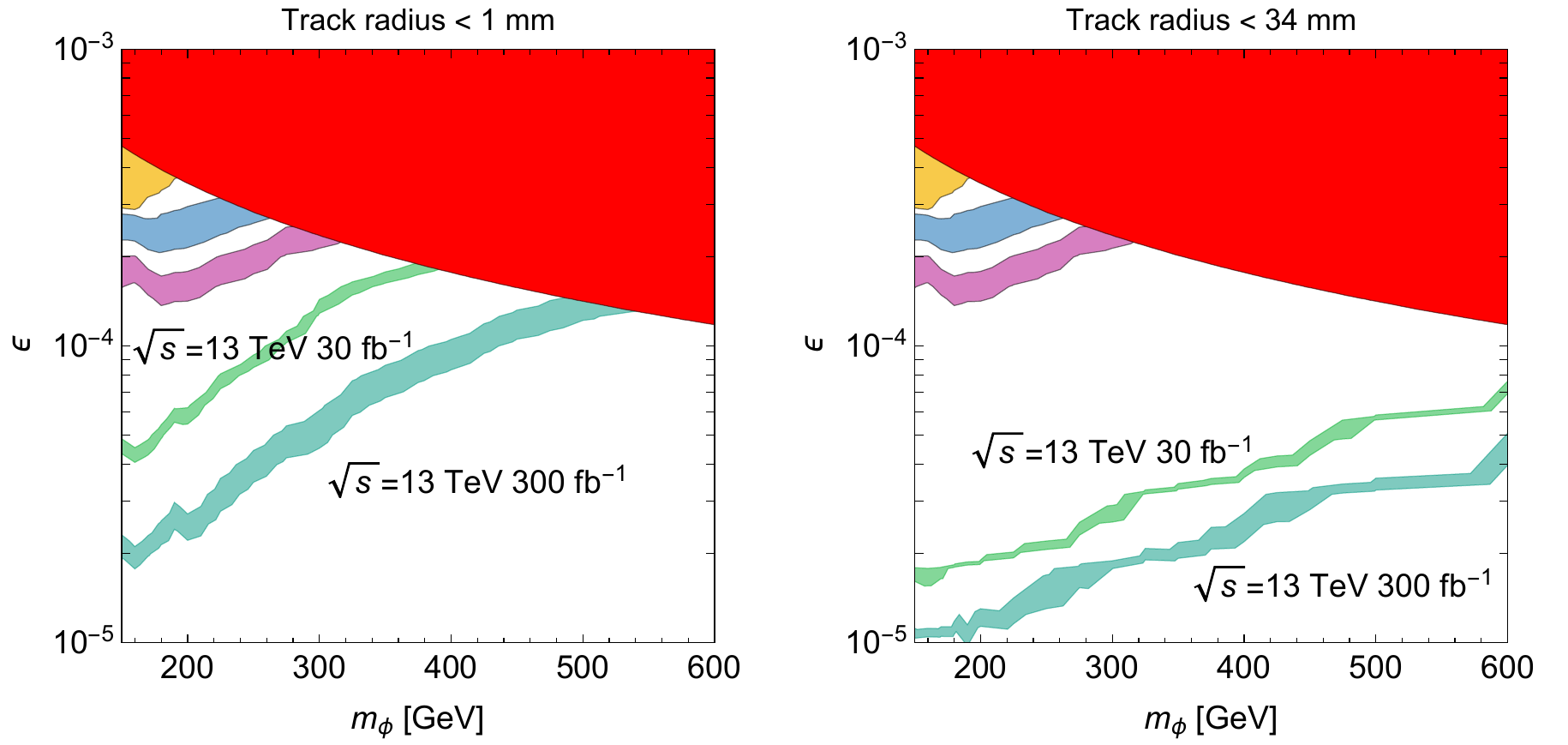}
\caption{Limits obtained when looking for events containing two lepton jets. The lepton jets are reconstructed using jet-substructure techniques and a BDT for classification. A lepton-jet object is required to have at least one track clustered into it. The left and right panels consider tracks which start within 1 or 34 mm of the primary vertex, respectively, which allows us to test for a larger range of $\epsilon$ values (due to the dark photon decay length). The yellow, blue, and purple regions are the areas which are, or can be, excluded using the reconstruction strategy of the original ATLAS search \cite{Aad:2015sms}, and the red region is excluded by the $T$ parameter constraints.  The green shaded bands are obtained by varying the expected background up and down by a factor of 2. The limit is expected to fall within the band and exclude everything above it.}
\label{fig:13TeV_withSubStruct}
\end{figure}

\fref{fig:13TeV_withSubStruct} shows the limits obtained using the boosted decision tree with the jet-substructure variables when requiring that there are two lepton jets in the event. This is therefore the same event topology as the ATLAS prompt lepton-jet search, but the lepton jets are defined with a different algorithm, allowing for more stringent cuts. The left panel uses the definition that the tracks must start within 1 mm of the primary vertex. Examining the green bands, it is clear that using these techniques results in improvements on the limits of up to an order of magnitude in $\epsilon$, and in the right panel, we see that the limits can be extended much further if the tracks can still be safely reconstructed to longer displacements. This greatly helps push the limits to lower values of $\epsilon$ where the decay lengths grow larger, and thus for the best sensitivity, it is important to allow the longest feasible track displacements.

As noted earlier, since \texttt{Delphes} ignores decays in the calorimeters, the lower $\epsilon$ points can have signal events where the calorimeters are missing energy deposition.  Given our search requires two lepton jets, this only affects dark Higgses of which the decay into two dark photons has one decaying early (satisfying the 34 mm requirement) and one late inside one of the calorimeters.  Since the calorimeters span a small range in radius ($\sim 1.15-4.25$ m), this only affects a fraction of these events.  Whether the limit would improve or worsen is, however, not clear.  Decays inside of the EM calorimeter could help, since it would increase the EM fraction, while decays in the hadronic calorimeter would decrease it.  The substructure variables on the other hand will benefit since they will be sensitive to the multiple prongs of the dark Higgs decays, which we have seen the BDT is efficient at picking up.  

\subsection{Observable aspects of the overarching model}

In the previous section, we improved the lepton-jet reconstruction, as well as the classification leading to substantial gains beyond the existing ATLAS prompt lepton-jet search strategy. Anticipating a future discovery of our lepton jets in our studied channel, the next important step would be to determine how the lepton jets are being produced and what the source of kinetic mixing is. For instance, as opposed to other models, where the dominant dark photon production mechanism is through a Higgs portal, our dark photons come directly from new scalar states. The presence of these states could be inferred in multiple ways, such as in $\phi^0$-decays to Higgs which could be looked for in events with lepton jets and a $b$ quark (or even $b\bar{b}$ resonances). A more detailed study of events containing lepton jets will help distinguish between different models of kinetic mixing.

\begin{figure}[t]
\includegraphics[width=0.8\linewidth]{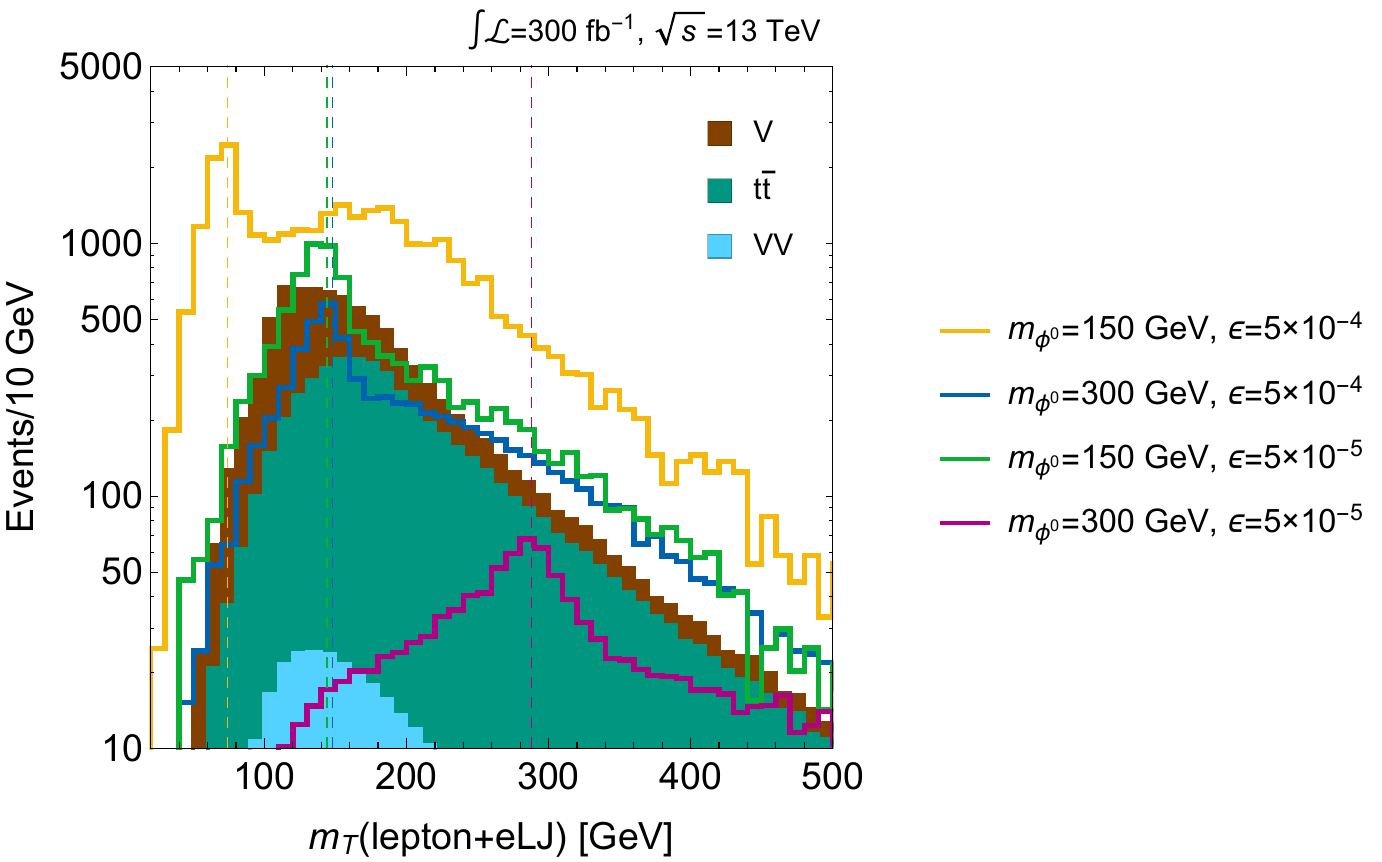}
\caption{Mass reconstruction for events with exactly one lepton and one electron lepton jet (using the 34 mm track displacement requirement). The dashed lines show the mass of the $\eta^{\pm}$ for each of the benchmark models. The $m_T$ spectrum shows edges near these masses. Note that the requirement of one lepton in the event makes the multijet background negligible.}
\label{fig:MT}
\end{figure}

Alternatively, most of the lepton jets we reconstruct come from decays of the lightest scalar, $\eta^{\pm} \rightarrow W^{\pm} + \text{lepton jet}$ (see Fig.~\ref{Fig:typical event}), where, in principle, the signal kinematics allow for a method of measuring its mass using the $m_T$ variable. To show how this would work, we examine events in which only one lepton jet passes through both reconstruction and classification (to avoid issues of combinatorics) and then select ones in which there is exactly one isolated, \texttt{Delphes}-reconstructed lepton. In Fig.~\ref{fig:MT}, using our 34 mm track displacement requirement, we have plotted the $m_T$ distribution obtained after combining the 4-momentum of the lepton jet and the lepton. The dashed lines show the mass value of the $\eta$ for the different benchmark models. In particular, the lighter models have a noticeable edge near the mass of the $\eta^{\pm}$ as expected from the $m_T$ variable, which suggests that information about the mass spectrum is accessible at Run 2 if the scalars are light enough.

Measuring the mass of $\phi$ is an important step in establishing that non-Abelian kinetic mixing is responsible for the lepton jets  and a concrete step toward measuring the parameters of the model.  In fact, almost all of this model's parameters are in principle measurable by upcoming  fixed-target and collider experiments.  Fixed-target experiments could measure the mass of the dark photon $m_{A_{D}}$ and kinetic mixing parameter $\epsilon$, giving two constraints on $g_D$, $\lambda_{\text{mix}}$, $v_D$, and $m_\phi$. From collider experiments, we can in principle determine all three scalar masses, which using \eref{eq:phi mass splitting} would provide two more constraints by measuring $m_{\phi}$ and  $\lambda_{\text{mix}}$, so with both types of experiments, we can measure all four parameters.   Finally, the most difficult parameter, $\kappa$, can be measured through the branching ratios of $\phi$.  Thus, if such a theory of dark photons is realized in nature, a combined experimental program could pin down the relevant theory parameters.

\section{Conclusion}
\label{sec:Conclusions}

The dark photon represents an interesting window into many classes of beyond the standard model physics, and when it is produced at a collider, it creates an entirely new object: the lepton jet.  Over the next few years, fixed-target experiments will be making an effort to discover this new particle, but the mass ranges they are probing kinematically disallow decays into muons, which are the most sensitive channel for colliders. Given the weak sensitivity of existing searches for electron-only lepton jets, this paper focused on introducing a new technique for these searches, in the hopes of being sensitive to a future dark photon discovery at the intensity frontier, and we have shown that there is a better way to find these elusive objects.

Based upon a recently proposed dark photon model with non-Abelian kinetic mixing \cite{Barello:2015bhq}, we examined the case where the dark photon is produced by scalars right around the electroweak scale: a mass range imminently accessible to the LHC. We examined constraints coming from Drell-Yan pair production of these new states by recasting existing ATLAS searches and showed that current constraints are weak, and, furthermore, that at 13 TeV, the LHC constraints are no longer background free and thus have limited gains with luminosity, which results in having sensitivity to less than half of the fixed-target parameter space. To overcome this, we modified an approach designed to find photon jets, proposed by Ellis \textit{et al.} \cite{Ellis:2012zp, Ellis:2012sd}, that relies on a boosted decision tree trained on jet-substructure variables.  We find that such an approach is highly successful for distinguishing electron lepton jets from hadronic jets, which allows a much larger part of parameter space to be searched for at Run 2 than previous methods.

This method is very successful in constraining the parameter space of the model, with the primary limitation being the displacement of the tracks, as decreasing the kinetic mixing strength drastically increases the dark photon decay length. We conservatively chose a maximum of $r=34$ mm for the initial appearance of the track to avoid complications arising from converted photons  and derived a limit which covers most of the fixed-target parameter space in $\epsilon$.  We also showed that once these events are seen many model parameters, specifically the scalar state masses, are obtainable with enough luminosity.

In the next few years, the LHC and fixed-target experiments have a tantalizing sensitivity to dark photons, especially if the mixing with the standard model is through the SU$(2)_L$ component of the photon. The methods we have introduced suggest that there is room for improvement in the LHC's approach to finding light dark photons, specifically in how analyses reconstruct and classify the lepton jets created by dark photon decays. By taking advantage of these improvements, we stand to learn much about dark photons and the sector in which they exist.


\section*{Acknowledgements}
We thank Stephanie Majewski, Tuhin Roy, Brian Shuve, David Strom, Eric Torrence, and Marat Freytsis for useful conversations. B.O. would like to express a special thanks to the Mainz Institute for Theoretical Physics for its hospitality and support. This work was supported in part by the U.S. Department of Energy under Grant No. DE-SC0011640. We also acknowledge 
the use of computational resources provided by the ACISS supercomputer at the 
University of Oregon, Grant No. OCI-0960354. 

\clearpage
\appendix
\section*{Appendix}

In this section, we used the ``Cuts" option of TMVA to see how a cut-based approach compares with the BDT. In this option, TMVA makes one rectangular cut (a maximum and a minimum) on each variable. A ROC curve is obtained by minimizing the background efficiency for different values of the signal efficiency. Table~\ref{tab_ROC} shows that the rectangular cuts perform $14\%$ worse than the BDT as evaluated by the area under the ROC curve for 1 mm tracks, while the 34 mm tracks are only $6\%$ worse.

\begin{table}[h]
\caption{Rectangular cuts compared to boosted the decision tree. The signal efficiency is evaluated at the BDT value or cut values which yield a background efficiency of 0.01.}
\label{tab_ROC}
\begin{tabular}{l | c c | c c}
\hline \hline
& \multicolumn{2}{c|}{ROC AUC} &\multicolumn{2}{c}{Signal efficiency}\\
\cline{2-5}
  & 1 mm & 34 mm & 1 mm & 34 mm \\
\hline
BDT & 0.912 & 0.961 & 0.37 & 0.66 \\
Rectangular cuts & 0.784 &0.902 & 0.33 & 0.64 \\
\hline
Percent difference & 14 & 6 & 10 & 3 \\
\hline \hline
\end{tabular}
\end{table}

While the ROC scores provide information about the signal and background acceptances over a range of working thresholds, we have chosen to work with a background acceptance of 0.01 in this work. Thus, the comparison relevant for the signal reach is the signal efficiency for a background acceptance of $1\%$. Table~\ref{tab_ROC} shows that the rectangular cuts are $10\%$ and $3\%$ worse for the 1 and 34 mm track length requirements, respectively. Thus, for roughly the same number of background events in our projections, the rectangular cuts would have worse exclusions because fewer signal events make it through.

\begin{center}
\begin{figure}[h]
\includegraphics[width=\columnwidth]{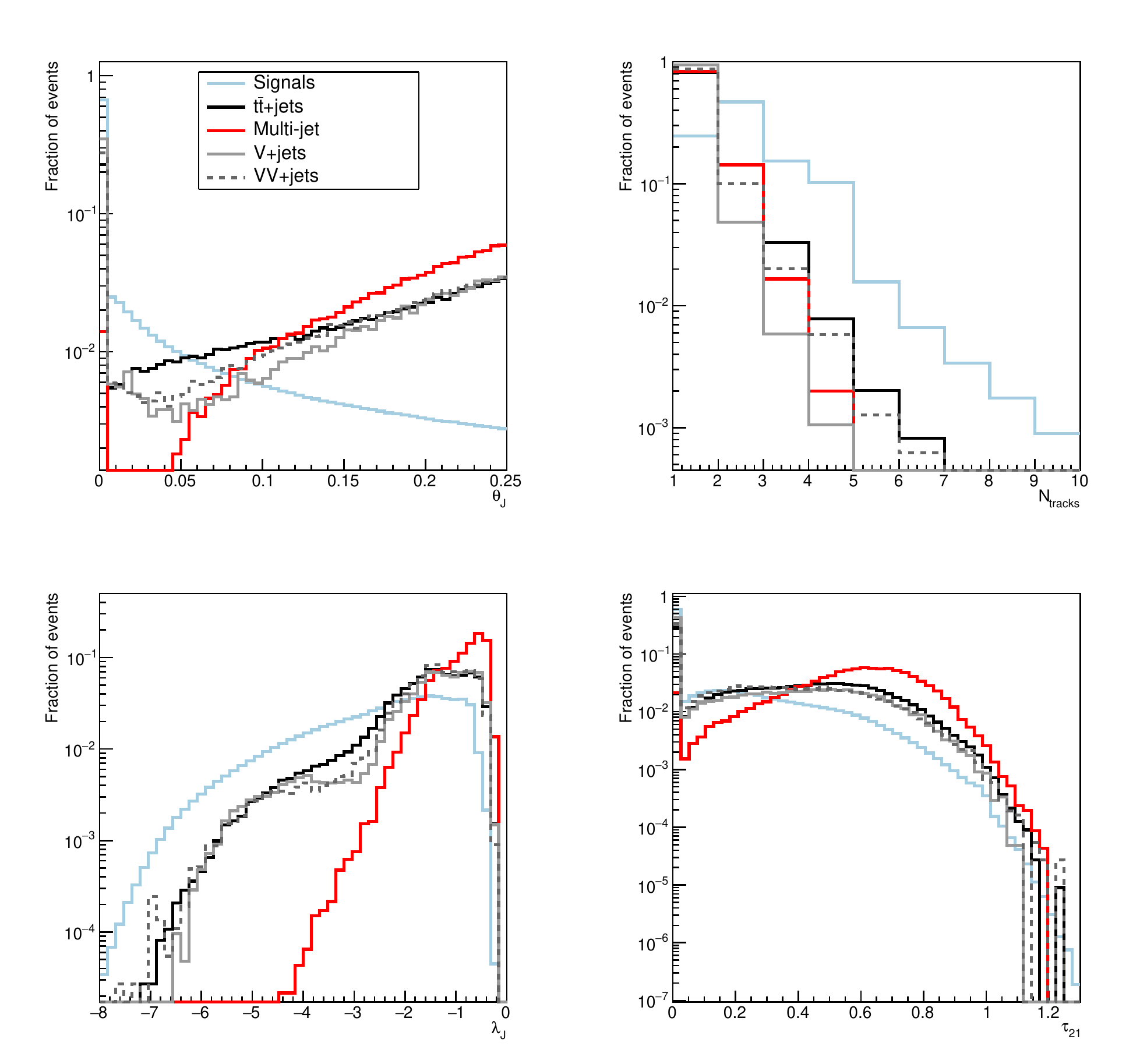}
\caption{Distributions for track length $< 34$ mm. We have already applied the preselection cuts of $\theta_J < 0.25$ and $N_{tracks} \ge1$. }
\label{fig:CutFlows}
\end{figure}
\end{center}

As a reference, Tables~\ref{tab_1mm} and \ref{tab_34mm} show the cuts used to obtain a background efficiency of 0.01, after the initial preselection. The distribution of the different variables considered are shown in Fig.~\ref{fig:CutFlows} for the 34 mm tracks. The 1 mm tracks distributions are very similar; however, the tightened requirements on the tracks also affects the preselection. Due to these tighter requirements, the best signal efficiency for a background acceptance of 0.01 is obtained without using any of the substructure variables, but only $\theta_J$ and $N_{track}$, as shown in Table~\ref{tab_1mm}.   Examining the cuts for $\lambda_J$ and $\tau_{21}$ for the 34 mm tracks in Table~\ref{tab_34mm} reveals that these cuts  are relatively weak and give much smaller background rejection,  when compared with $\theta_J$ and $N_{track}$.  This is confirmed by the plots of Fig.~\ref{fig:CutFlows}, which shows that cutting hard on these variables would lead to more background rejection at the expense of signal efficiency.      Another compelling point is that the final comparison to the BDT in each table shows that for a fixed overall background rejection the BDT rejects the multijet sample at the expense of worse rejection of the vector boson backgrounds. 

\begin{table}[h]
\caption{Cut flow for 1 mm tracks to achieve a background efficiency (after preselection) of 0.01. The best rectangular cuts only use $\theta_J$ and $N_{track}$ for this value of the background efficiency.}
\label{tab_1mm}
\begin{tabular}{c | c c c c}
\hline \hline
Cut &  $t\overline{t}$+jets& Multijet & V+jets & VV+jets \\
\hline
Preselection & $1.13\times10^{-2}$  & $5.51\times10^{-3}$ & $8.56\times10^{-3} $ & $1.27\times10^{-2}$ \\
$\theta_J < 0.0720$ & $4.50\times10^{-3}$ & $1.60\times10^{-4}$ & $3.77\times10^{-3}$ & $4.64\times 10^{-3}$\\
$N_{track} \ge 2 $ & $1.50\times10^{-4}$ & $1.05\times10^{-5}$ & $1.15\times10^{-5}$ & $9.50\times10^{-5}$\\
\hline
BDT Efficiency & $1.49\times10^{-4}$ & $6.68\times10^{-6}$ & $1.94\times10^{-5}$ & $1.51\times10^{-4}$ \\
\hline \hline
\end{tabular}
\end{table}

It is interesting to see that the variables with the strongest separating power are nearly the same as already used by ATLAS.
The biggest difference between our method and the ATLAS search is in the reconstruction of the lepton-ket object. ATLAS's lepton-jet objects are formed out of clustered tracks and are required to have two tracks with $p_T > 10\gev$, with one tagged as an electron. In contrast to this, our lepton jets are made from clustered tower hits (in the ECAL and HCAL), and then we count how many tracks get clustered into the jet object. This difference in reconstruction philosophy (building out of tracks or energy deposits) affects both the reconstruction efficiency as well as the number of tracks associated with a lepton-jet.  These differences are relevant enough to make a substantial improvement to electron lepton jet efficiencies with substantially higher background rejection.

\begin{table}[t]
\caption{Cut flow for 34 mm tracks to achieve a background efficiency (after preselection) of 0.01.}
\label{tab_34mm}
\begin{tabular}{c | c c c c}
\hline \hline
Cut &  $t\overline{t}$+jets & Multi-jet & V+jets & VV+jets \\
\hline
Preselection & $1.47\times10^{-2}$  & $5.95\times10^{-3}$ & $9.59\times10^{-3} $ & $1.42\times10^{-2}$ \\
$\theta_J < 0.0606$ & $4.61\times10^{-3}$ & $1.53\times10^{-4}$ & $3.82\times10^{-3}$ & $4.72\times 10^{-3}$\\
$N_{track} \ge 2 $ & $1.95\times10^{-4}$ & $1.39\times10^{-5}$ & $1.15\times10^{-5}$ & $1.00\times10^{-4}$\\
$\lambda_J < -0.429$ & $ 1.94\times 10^{-4} $ & $1.31 \times10^{-5}$ & $1.11 \times 10^{-5} $ &  $1.00\times10^{-4}$\\
$\tau_{21} < 0.989$ & $1.93\times10^{-4}$ & $1.31 \times10^{-5}$ & $1.11 \times 10^{-5} $ &  $1.00\times10^{-4}$\\
\hline
BDT efficiency & $1.91\times10^{-4}$ & $7.32\times10^{-6}$ & $2.72\times10^{-5}$ & $1.15\times10^{-4}$ \\
\hline \hline
\end{tabular}
\end{table}

\end{spacing}
\begin{spacing}{1.1}
\clearpage
\bibliography{LHC_NAKM}
\bibliographystyle{utphys}
\end{spacing}
\end{document}